\documentclass{article}[11pt]
 
\newcommand {\bea}{\begin{eqnarray}}
\newcommand {\eea}{\end{eqnarray}}
\newcommand {\be}{\begin{equation}}
\newcommand {\ee}{\end{equation}}
\newcommand {\bc}{\begin{center}}
\newcommand {\ec}{\end{center}}

\def\lsim{\mathrel{\rlap{\lower4pt\hbox{$\sim$}}
    \raise1pt\hbox{$<$}}}               
\def\gsim{\mathrel{\rlap{\lower4pt\hbox{$\sim$}}
    \raise1pt\hbox{$>$}}} 

\usepackage{amsmath, amssymb, amsfonts, amsthm}
\usepackage{stmaryrd}
\usepackage{graphicx}
\usepackage[all]{xy}
\usepackage{color}
\usepackage{bm}
\usepackage{subfigure}
\usepackage{graphicx}
\usepackage{mathabx}
\usepackage{multirow}
\usepackage{setspace}
\usepackage{epstopdf}
\usepackage{makeidx}
\usepackage{multicol}  
\usepackage{braket} 
\usepackage{geometry}
\graphicspath{{./Pictures/}}

\author{Clifford E Chafin\\\ \small{Department of Physics, North Carolina State University, Raleigh, NC 27695} \thanks{cechafin@ncsu.edu}}
\title{The Quantum State of Classical Matter II: Thermodynamic Equilibrium and Hydrodynamics}

\usepackage[pdftex, pdfusetitle, plainpages=false, 
				letterpaper, bookmarks, bookmarksnumbered,
				colorlinks, linkcolor=black, citecolor=black,
	         filecolor=black, urlcolor=black]
				{hyperref}

\begin{document}
\begingroup
\let\clearpage\relax
\maketitle
\begin{abstract}
In the previous companion paper, we proposed a subclass of wavefunctions to describe macroscopic solids that resolved and extended the theory quantum measurement and gave a more specific treatment of quasiparticles.  Here we extend these notions to thermalization of solids and gases and to gas state hydrodynamics.  This gives a modification of the thermodynamic limit to justify the canonical averages for ``typical wavefunctions'' without the use of ensembles.  The energetic cost of vorticity is contrasted in the classical and ultracold gas limits.  From this perspective, we then examine the applicability of thermo and hydro to ultracold gases and compare with the implications of pure state evolution.  We illustrate how the proposed quantum limits on viscosity could be reinterpreted in terms of Schr\"{o}dinger induced evolution of the one-body density function but some history dependent measurable properties should still persist.  
\end{abstract}
\section{Introduction}

In the previous article we discussed the necessary constraints on a many body wavefunction to correspond to a ``classical'' solid, specifically, one describable by a three dimensional set of variables governing its shape, composition and dynamics.  These were far-from-eigenstate wavefunctions that exhibited a long lasting quasi-ground state and quasi-eigenstates corresponding to phonons.  While the origin of such states is still somewhat unclear, the specificity of this description was robust enough to explain, from a quantum mechanical point of view, several points including: classical motion, the distinction of phonons as quasiparticles that carry no true momentum, some subtleties for bodies with discrete symmetries, and even a natural explanation and extension of quantum measurement.  

Gases introduce some new complications.  The long lasting localization macroscopic bodies can possess that allows classical descriptors including well defined shape and orientation is not present for them.  The hydrodynamic and thermodynamic properties of these are typically inseparable.  Our task is to first provide a single wavefunction understanding of thermalization and overcome the long standing apparent contradictions of quantum statistical mechanics with quantum evolution itself.  We will show that the meaning of the ensemble averages can be reinterpreted to be consistent with the quasi-local properties of a single wavefunction that gives the complete description of the system at all times.  Temperature can be defined in a satisfactory manner but we will have to reconsider the thermodynamic limit as more involved than just involving large numbers of particles.  Since quantum evolution still holds we don't get equilibration for eigenstates and few state superpositions of them.  Thermodynamics thus depends on having a sufficiently broad distribution of such eigenstates.  

Our motivation for such specific descriptions of classical matter in terms of single wavefunctions is to compare with more evidently quantum ones.  If a single descriptor and evolution equation can describe measurement and thermodynamics of apparently classical system then it will have stronger credibility in its implications for ultracold gases.  The state of the field of many body quantum mechanics is ``ansatz-heavy.''  When the bar for conceptual justification of a calculation is low, methods proliferate and it is tempting to assume our way to standard classical form of motion from which such methods provide thermodynamic variables and transport coefficients.  Even when these results turn out to be valid, we lack sufficient understanding to jusficy them over other less successful attempts.  

Hydrodynamics introduces several more problems.  Here we must give an explanation for the apparently greatly reduced set of degrees of freedom a classical gas exhibits versus quantum many body wavefunction.  We demonstrate that there are many dynamic nonclassical gas solutions and that bulk expansion, generally associated with bulk viscosity, can pick up nonclassical nonhydrodynamic behavior.  Nevertheless, for a physically recognizable set of states that correspond to hydrodynamics and some not-to-violent motions, the classical hydrodynamics of gases is recovered.  The primary difference is that corrections to these can involve deviations from the ``classicality'' of the system and its representation in terms of such a limited 3D variable set.  

Ultracold quantum gases will be shown to often not have well defined thermodynamics or hydrodynamics although static behavior and small oscillations near the ground state can be well modeled with the Euler equation.  These systems do, however, relax in the limited set of variables we typically probe: the one-body density function.  The extent to which this can be associated with a viscosity, a purely hydrodynamic concept, is discussed.  
Experiments are proposed to reveal history dependent effects hidden in the relaxed states of these clouds.  

The origin of superfluidity and a derivation of the two-fluid model are still waiting.  The theory of ultracold gases often more embraces the two-fluid model and its concepts as building blocks rather than using them as an opportunity to test its limits and ultimately give a justification for it.  This is undoubtable a very hard problem to treat from the other direction.  Classical hydro is expected to govern the normal component in the two-fluid model and we are still lacking a good quantum and microscopic understanding of classical liquids.  Some agreement has been found by combining molecular dynamics simulations and linear response theory \cite{Eu} but it is still unclear how much of a fundamental understanding will follow from this approach.  The biggest motivation for the following is to elucidate classical and extreme quantum systems from a single description in hopes of finding the right questions, hence the right calculations, to explain such persistent phenomena.  A better understanding of ultracold gases is certainly a positive step in this direction.  

\section{Thermalization}

\subsection{Conceptual Problems }
Thermalization is most natural as a classical notion.  It implies that observable history dependent effects are obliterated over time so that a system with fixed energy (and other conserved quantities) tends to states that are apparently equivalent.  This seems to violate the time reversibility of the fundamental equations of motion.  Usually this is phrased in terms of the ``microscopic'' equations.  In the classical case this is resovable.  Larger scale motions and less stable small scale ones are generally short lived over the longest time scales \cite{Jeans}.  This does not mean they cannot return arbitrarily close to their original state.  In fact, for isolated systems of fixed size, the Poincare cycle implies they must and extensions of this result show they must do so with arbitrarily long arithmetic progressions \cite{Petersen}.  However, the long time behavior is dominated, in almost all cases, by global uniformity of motion and a local particular velocity distribution that is well attained by averaging over short time scales of the system.  If there is a conceptual problem, it is how such initial data arose that led to such an apparently unidirectional time.  The anthropomorphic principle gives a suitable resolution.  If there were no such gradients present, there would be nothing to power living beings to observe it.  

Quantum systems are more troubling for a number of reasons.  It is standard to invoke a kind of loose standard to quantum arguments that allows a convenient mixing in of classical notions.  When forced to be more specific about this, there are numerous problems.  If we wish to think of quantum effects as driving the ``microscopic'' equations, we have the situation that quantum effects can manifest on large scales as in the case of superfluidity and superconductivity.  In the case of ultracold gases, we will show there can be other large scale quantum effects that do not vanish as well.  In the previous sections \cite{Chafin} we saw reason to not exclude macroscopic superpositions of a certain class.  This makes it unclear if we can expect our observations to settle down into an apparent uniformity and consistency of motion with initial larger scale variations vanishing into unobservable smaller scale ones.  More frustrating is that quantum dynamics never change the eigenstate distribution of a given state so the origin of ``equilibration'' is even more mysterious.  

Classical gases have the property of thermalization.  
This is most easy to see in the case of a gas where a system with a spatial energy and momentum distribution that is ``uniform'' in some spatial average over mean free path size parcels 
equilibrates to the Maxwell-Boltzmann distribution; usually in just a few collision times.  Analyses of this state can be from the classical kinetics \cite{Boltzmann} of billiard-like motions to more abstract formulations in terms of phase space distributions, ergodicity or the partition function \cite{Tolman}.  The former treatment is the most direct, though much more demanding.  It has the virtue of being a \textit{dynamic} approach rather than a kinematic one based on more abstract notions (that we would ultimately need to validate by dynamic means) and therefore include information on fluctuations and transient relaxation.    
It is easily seen that the delocalization rate of gas particles at usual densities causes the wavefunction spreading to rapidly exceed the interparticle separation.  This weakens the conceptual link of this model with a realistic gas that we expect to be described by a wavefunction.  We will pursue this when we discuss hydrodynamics.  

Phase space is well defined for the classical case as any set of N particles can be represented as a point in 6N-D phase space.  Allowing for some uncertainty in the initial data, we may define a weighted volume in this space and evolve it accordingly.   
Liouville's theorem guarantees the volumes of parcel are preserved and so suggests that the canonical momentum gives the natural measure on this space.  Most physical Hamiltonians give strong stretching and folding of this volume in the manner of a Smale Horseshoe map\cite{Abraham}.  In many cases this mapping is ergodic; specifically the time average at each point tends to be uniform in phase space.  This lead to the temptation to use such phase space averages to give thermodynamic averages.  
However, not all systems display ergodicity and, even for the ones that do, the sampling time to fill out the phase space to a given level of global uniformity is generally far greater than any observation time of the system.  Boltzmann understood that ergodicity was a poor foundation for statistical averages and 
was determined not to involve ergodicity in the definition of thermal equilibrium values.  The long lasting confusion that ergodicity was essential can be traced to an early and well read article by the the evidently confused P. and T. Ehrenfest \cite{Boltzmann}.  
Additionally, there are other conservation laws to preserve.  This demonstrates to us that to discuss thermal states we really need a ``typical'' element of our system that gives the averaged thermodynamic quantities as nearly-always holding conditions of the system rather than the sum over an ensemble (though we expect there to be generally no difference in the resulting values).  

In quantum mechanics, the partition function and thermodynamic averages have greater problems and a long frustrating history.  The Boltzmann factor is so strongly validated by experiment that theory must offer an explanation.  However, the justification in quantum treatments of statistical mechanics is even worse than in the classical case.  One root of the problem is that quantum systems do not ``equilibrate.''  Eigenstates and finite superpositions have predictable stationary or periodic behavior.  Infinite state superpositions are not generally periodic but still maintain a fixed ratio of occupancy.  If ``equilibration'' depends on arriving at a special subset of distributions of states, this never happens.  

The microcanonical ensemble \cite{Tolman}\cite{Landau:smI} is of fundamental importance since it is the starting point to later generate the more practical canonical and grand canonical ones.  
To form the microcanonical ensemble one takes a sum over the \textit{eigenstates} of energy $E_{i}$ within some $\delta E$ spread and, for large enough particle number $N$, constructs a quasicontinuum of such states with the number of such states as $\Omega(E)\equiv e^{S(E)}$.  Weakly coupling with an another much larger such system allows the subsystem's eigenstates to remain near eigenstates with long lasting validity and products of such functions give good approximations to the net system's eigenstates.  The exponential growth of the density of states and an enumeration argument shows that the most likely ones are from a sharply peaked distribution where the energy per particle in both the system+reservoir and the system individually is $\epsilon=E_{Net}/N$.  Defining the temperature as $T^{-1}=dS/dE$ for this closed system\footnote{Volume, net energy and particle number are fixed and T is explicity a function of them.} we arrive at the ``probability'' of a state with energy $E_{i}$ in this distribution of as $p_{i}\sim e^{-E_{i}/T}$.  Combining this decreasing probability with the exponentially increasing number of states $\Omega(E)$ gives back our sharply peaked distribution for our system about $E=N_{sys}\epsilon$.  

%The one body observables my exhibit long return times.  

Of course, there is no reason to assume that the system is in an eigenstate and there are many superpositions that have net energy $E$ that do not come from the narrow band of many body eigenstates near this value.  Since any weighted combination of eigenstates are allowed and this never changes, it is hard to see how this ``probability'' has any meaning as a likelihood from an equilibration.  We can imaging scenarios where we randomly turn on and off coupling to the reservoir or otherwise perturb the system to try to bring this situation about but classical systems seem to reach equilibration in isolation.  A famous method for treating quantum equilibration by coupling to a ``bath'' of harmonic oscillators is the Caldeira-Leggett model \cite{Caldeira} yet each of these seems like just a more elaborate contrivance to obtain the Boltzmann factor we know is somehow important.  

Energy is not the only conserved quantity in physics even though it is the one that plays the singular role in almost all statistical mechanical calculations.  Galilean translations alone give ten  global invariants.  A rotationally invariant potential gives angular momentum conservation.  In the classical case, we can just impose this as another subfoliation of our fixed energy surfaces in phase space.  (It is interesting that this is typically ignored for net $L=0$ systems yet we still tend to get correct results.)  Nonrotationally invariant systems can give nonzero time averaged $L$ conservation for many solutions as in the case of the vertical pendulum so this is a concern for thermodynamics even when rotational invariance fails.  In the quantum case, it is unclear what to do.  Do we restrict our microcanonical ensemble to only consider eigenstates with the desired angular momentum?  Do we bias the weights in the distribution in some yet unspecified way?  Instead of the ensemble, should we just look at the totality of possible wavefunctions with fixed $L$ and $E$ and average desired quantities?  If so, will this agree with the microcanonical result as $L\rightarrow0$?  The canonical ensemble does provide an ansatz for handling this with the thermodynamic average, specifically $\Braket{|Q-L\omega|}$, but this seems like one more bit of subject lore to be remembered rather than explained.    

There is no natural analog of phase space in quantum mechanics.  Classical systems are described by position and momentum but quantum ones are described by vectors in Hilbert space or, more humbly, smooth bounded wavefunctions where the value at each point\footnote{Here we allow ``point'' to be both a postion in coordinate and spin space.} is a complex value.  There have been arguments using a vague partitioning of the $(p^{3N},q^{3N})$ phase space via the uncertainty relation.  Some of these have led to useful results but it is unclear how to make these precise and why they should be meaningful.  

The canonical ensemble introduces ``mixed'' states that do not correspond to any eigenstate of the system \cite{Neumann}.  Such objects are inconsistent with the evolution of a single wavefunction.  One can attempt to justify such collections by assuming these other states simply represent some uncertainty in our knowledge as observers outside the system.  In this case, it is hard to see why the equilibration and averages should ultimately depend on them and why further measurements should not restrict the set to a more refined subset of such a distribution.  If one believes the many body quantum objects we study in cold gas traps are truly just evolving wavefunctions, it is hard to see how one could justify using the canonical or grand canonical ensemble for results beyond what would automatically hold by consistency with the microcanonical one.  Furthermore, why should such objects even be limited to such a narrow range of energy eigenstates in its construction and how does introducing angular momentum to the initial data alter this? 

In the case of typical macroscopic objects, we have come to expect some sort of disconnect between the worlds of quantum and classical dynamics.  As such, we get more comfortable with using quantum reasoning to determine microscopic properties that map on to the macroscopically classical physics.  In the case of effective field theory and green's function approaches to transport in hydrodynamics, we assume that classical hydro holds and then introduce linear response theory to compute the relevant parameters in the equations.  In a sense we have assumed our way almost to the conclusion and derived the missing values.  
From such a picture, it is not such a great step to then introduce some additional baggage of the mixed states.  However, for our cold gas traps we are more directly confronted with the inconsistency.  A trap is with one particle is clearly a pure state.  So is one with two, three and so on.  At what particle number do we transition to a mixed state?  Is this Bayesian uncertainty in our knowledge a result of large N? coupling to the outside world?  Is it just a harmless tool that works because of sharply peaked distributions?  

The difference between quantum and classical statistical mechanics can be summed up this way.  A classical gas at fixed energy will almost always be in a state that is locally reasonably described as thermal.  An isolated quantum gas is in whatever distribution of eigenstates it starts in and this never changes.  Classical gases rapidly tend to a well defined hydrodynamic flow but in the  quantum case, the space of configurations is so much larger that, as in the case of solids above, a well defined velocity profile $v(x)$ seems like a rather special subset of them.  The evolution of such a general state will be considered when we discuss hydrodynamics.  
%\footnote{The resolution for the thermal properties we suggest below is that, generally, the particular distribution of eigenstates is irrelevant when it comes to the local distribution of currents, at least when the energy density is large enough relative to the number density.  The measure of this is better done through a ``nearly free'' basis associated with an optimal quasiparticle spectrum.  Since this basis evolves through a different Hamiltonian we find that energy distribution of the true states can indeed change relative to it.  }

%We previously found a justification for describing classical condensed matter as a subset of far-from-eigenstate wavefunctions.  Synchronously with this, we expect the thermal excitations of it to be similarly limited in generality.  Beyond justifying thermodynamics with a typical set of wavefunctions for macroscopic matter and finding the extent of their role in smaller and cold systems,  we will examine the question of the apparent temporal asymmetry of such systems.  
%The Poincare cycle shows that a system with a compact phase space returns arbitrarily closely to its starting position repeatedly.  This is in complete contradiction with experience and the second law.  The return time is typically much longer than observable but such considerations are related to short time and small scale fluctuations in entropy \cite{Jarzynski}.   

In the spirit of seeking subset of wavefunctions for classical matter that corresponds to ``thermal'' states we will begin by giving a criterion for the ``thermodynamic limit.''  Traditionally, this is the $N\rightarrow \infty$ limit.  For trapped gases and small clusters, this seems overly restrictive since some sort of equilibration can occur for them for very modestly sized $N$.  Furthermore, we will show that there is both experimental and theoretical reasons to believe that isolated cold atomic clouds can persist in very far from what we would consider a thermal state.  These can have exotic momentum distributions and lack the properties of self-thermalization and thermalization with other bodies they are made to interact with.  

%Justify and give meaning to the ensemble averages.  Give thermo limit in terms of mfp and N.  When L ne 0 show there is another scale that allows quantum vorticity effects in large bodies.  

There has been recent work utilizing the properties of very high dimensional spheres (Levy's theorem) \cite{Popescu} to show that in the high dimensionality of a many body system, the thermal distribution is overwhelmingly favored.  This is kinematic argument and gives no notion of how such states would be arrived at.  Other methods to attempt to justify the thermodynamic dominance of the microcanonical states generally involve some Hilbert space measure \cite{Brody} \cite{Goldstein}  and a coupling to an external pure state universe.  External coupling is known not to be important for equilibration and, moreover, there is no immediately obviously physical measure on this space.  Such a natural measure could only be defined by the dynamics of the system in choosing it in equilibration, not the mathematical aesthetics of a theorist. These methods typically don't include any other conserved quantities but energy that the system must conserve.  Ultimately, we would like dynamic understanding of equilibration that incorporates all of them.  %Their result seems to be insufficient to understand how ``nonequilibrium'' systems arrive at equilibrium themselves and is likely inconsistent with such a process.  

%Based on the usual consideration of high dimensional states of the microcanonical ensemble we expect the sum for $\Psi_{th}$ to be dominated by a narrow shell of states in $\mathbb{R}^{3N}$ at radius $|k|=\sqrt{2mE}/\hbar$.  Methods to attempt to justify the thermodynamic dominance of these states generally involve some Hilbert space measure \cite{Popescu}\cite{Brody} \cite{Goldstein}  and a coupling to an external pure state universe.  Unfortunately, there is no immediately obviously physical measure on this space.  Such a natural measure could only be defined by the dynamics of the system in choosing it in equilibration, not the mathematical aesthetics of a theorist.  Furthermore, the true eigenstates of the system are difficult to represent when interactions are present and have the frustrating property of never equilibrating!  By asserting that a subset of wavefunctions are typical both for thermal and ground states of macroscopic matter we avoid this.  Self consistency then requires that such states be preserved but no notion of their commonality is required.  Even the NFA basis is just a convenient description to capture the properties of such dominate states.  It is unlikely that any of the true eigenstates can be represented in the narrow shell of states energetically near it.  

We argue that wavefunctions of systems that start off sufficiently ``broadband,'' not made of a very small $\Delta E$ spread or number of energy eigenstates and for which the expected standing \textit{and} dynamic fluctuation wavelengths are much smaller than the interparticle separation, tend to long time stable distributions of local current flux balanced systems.  This is generally independent of the details of the energy distribution of the superposition and gives a new way to assign meaning to the ``ensemble averages'' in terms of a typical \textit{single} wavefunction.  Local dynamics of such a wavefunction near the scattering centers naturally leads to equipartition and gives a reason for the success of classical kinetic results of thermodynamics.  

The ``thermal wavelength,'' $\lambda_{th}=\frac{\hbar}{2\pi m k_{B} T}$ is a basic parameter used the thermodynamic discussions of quantum systems.  The arguments using this are typically vague but this parameter does give a unique length scale in terms of the temperature.  We will be primarily interested in wavefunctions far from eigenstates that approach a kind of local regularity in their behavior independent of the particular distribution of eigenstates they are constructed from.  For fermions with an energy per particle $\epsilon\lesssim E_{F}$ we have a lot of curvature to the wavefunction we don't want to consider as ``thermal.''  The traveling component of the waves associated with local currents is the part that we are concerned with.  In some cases there may be states with too little energy above the ground state to give what we consider thermal, specifically a local kind of universality independent of the initial data.  

It is convenient to have a term to describe the kinds of wavelengths that appear in the currents in both of these cases.  Assuming some dominant local frequency exists at a point, we can express $\lambda_{typ}$ in terms of the local current $\tilde{J}=-\hbar\rho \frac{1}{m_{i}}{\nabla_{i}}\Phi$ where tilde's indicate a many body (3N component) vector and $\Phi$ is the phase of the many body wavefunction.  In a region with nearly constant potential height, the nonzero current part of the oscillations have wavelength (along the $x_{i}$ coordinate direction) $\lambda_{typ}^{(i)}=2\pi/|\nabla_{i}\Phi|$. 
The updated ``thermodynamic limit'' for quantum systems will then be $N\rightarrow\infty$ \textit{and} $\lambda_{typ}\ll d_{mfp}$ where $\lambda_{typ}$ will correspond to the thermal wavelength in the limit of high energy ($E\gg E_{F}$ or $E\gg E_{gs}$ for fermionic and bosonic atoms respectively).\footnote{The thermal state of metal band electrons seems like it would be a counterexample since the thermal oscillations in wavelength can be much longer than the mfp.  However, the net effect will still usually be governed by these atomic wavelengths since the net wavefunction is a function of cores and electrons and the cores typically give much shorter wavelength.  The case of ultracold fermionic gases is different since the cores themselves typically have very short wavelength.}  

For later reference, we extend the above notion more precisely and notice that the kinetic energy of a wavefunction can be decomposed uniquely into standing and traveling wave components.  This is convenient when we wish to discuss the kinetic energy that arises from vorticity (hence angular momentum) and currents from thermal motion independently of the, sometimes dominant, standing wave component especially at very low energies.  
For the one-body case, $\Psi=A(x) e^{i\phi(x)}$.  The kinetic energy density of the wavefunction decomposes as $\mathcal{E}=\frac{\hbar^{2}}{2m}A^{2}(- A''/A+\phi'^{2})=\mathcal{E}_{s} +\mathcal{E}_{j}$ for the static and current components.  This obviously generalizes for many body wavefunctions.  We thus have a local decomposition of energy density $E=E_{s}+E_{j}+U$.  Extensions of virial results are possible.  For example, classical gases have $U\approx0$ and $E_{s}=E_{j}$.  Eigenstates (with Dirichlet boundary conditions) in potentials without rotational symmetry give $E_{j}=0$.  The energy of solids are dominated by $U$.  If we let $\Delta U$ be the difference in potential energy over the ground state, harmonic solids also have $E_{j}+\Delta E_{s}=\Delta U$, by the usual virial theorem.    

%The fermion dof also obey Boltzmann probabilities.  The dos is what changes.  

%The photon coordinates can slice the function as well.  

%What is the effect of internal radiation fields? ***

%having a large distribution of frequencies and that have typical wavelengths much smaller than the interparticle separation and cross sections, tend to a particular class of states that we recognize as thermalizing both internally, and with external such ones.  The number of particles turns out to be not very important but does play a role in the size of the fluctuations observed.  A self consistent transitive definition of temperature will follow that is related to a particular density of quasi-stationary states even though the system will never be close to one them.  

%Give a condition on the validity of thermodynamic states.  

\subsection{Typical Thermal States}
Let us consider some first guesses at a typical thermal wavefunction and see why they don't work.  The microcanonical distribution suggests that we choose a wavefunction from the span of eigenstates sufficiently close to $E_{0}$.  If we make this width narrow enough we eliminate all temporal fluctuations.  Spatial correlations will generally persist and this is what is generally referred to as the quantum fluctuations of a system.  Temporal fluctuations are often treated by imposing stochastic forces but, presumably, a good approximation to the actual wavefunction would let us read these off directly. 
Since we know that temporal fluctuations are an important part of thermodynamics, we do not use the microcanonical ensemble as our starting point.  

%***The system can be assigned local values 3D when it is a near a product function.  If it evolves so this persists we can develop a 3D hydrodynamics.  Gas expansion may violate this.  
%  
%  ***Compare the case of a bound gas in a trap.  Classically, this is never stable until N is so small that there is not enough KE total for even one particle to escape.  Quantum case allows a finite fraction of ess to be bound.  Consider thermal expansion of traps that are anisotropic releases.  Slow deformations of trap alters the ess in a reversible manner.  What is the change in the E, r distributions?  If the iteration always uses the same ones do we get a persistent quadrupole bias in the result?  
%  
%  ***the 3D local case allows the NFA basis to be exact on small scales.  GF scattering methods for damping and transport are fine when system is globally 3D.  
%  
%  ***In general will NFA basis give equivalent averages and flucts as typ psi in time?  What is equilibration rate?  Will a very broad bimodal distribution equilibrate?  Very high and low freqs may not be able to locally scatter to MB.  Evolve the true psi on NFA basis and throw out phase.  Master equation?  
  
 %This would also be a promising place to start on extending the recent and exciting work on short time entropy fluctuations \cite{Cohen}\cite{Jarzynski} that, to date, have had only classical discussions.  

The next logical choice is to try the canonical distribution as a guide.  Noting that $P(E)=Cg(E)e^{-\beta E}$ where $g(E)$ is the (many body) density of states, 
%and we have assumed that no degeneracy is important at this value of $\beta$, 
we try a trial wavefunction
\begin{align}
\label{trial}
\Psi=\sum_{k} g(E_{k}) e^{-\beta E_{k}} e^{i\theta(E_{k})}\Psi(E_{k})e^{iE_{k}t/\hbar}
\end{align}
There is quasicontinuum of energy levels and we choose one representative from each level $E_{k}$ with a random phase specified by $\theta(E_{k})$.  It is not clear how well defined this definition is since we can keep refining our $\delta E$ separation of levels until the quasicontinuum approximation fails and eventually we include every such state.  Spatial coordinates are suppressed but the time evolution is explicitly included.  Two major problems exist with this wavefunction.  First it is unclear how a system ever arrives at it under (reversible) Schr\"{o}dinger evolution.  Second, some of the ``representative'' states of each energy level can be highly anisotropic both in their 3D projections and in their many body directions e.g. the high angular momentum eigenstates and the strongly interacting ultracold gas ground state.  Experience with high temperature gases tells us that isotropy is favored and correlations disappear.  (In terms of a wavefunction this would imply it averages to be isotropic on scales larger than the mean free path\footnote{The meaning of ``mean free path'' in the quantum case should be clarified.  For the case of short wavelengths compared to the particle cross sections, we have geometric scattering and the meaning is obvious.  This is primarily the case where we argue thermalization occurs.  For phonons, which don't have an easy packet construction, as argued earlier, so the collision time is a better measure.  It can be defined by the time our quasiparticle picture maintains its product function like validity.} 
 when rotated about hyperangular directions.)  It is probably true that a random wavefunction from the space spanned by this energy width $\delta E$ has these properties.  However, the previous reasons make it clear we should work harder at finding what a typical thermal wavefunction should look like. 
Let us begin with a proposition that describes what equilibration means for a wavefunction. 

{\bf Equilibration Condition }: The scattering rates of many body currents at each frequency from scattering centers about the two body diagonals along each single body coordinate direction must balance.  Specifically, 
\begin{equation}
\Braket{|\frac{1}{m_{i}}\Im\nabla_{x_{i}}\Psi(\tilde{X})|_{x_{i}\approx x_{j}}|}=\Braket{|\frac{1}{m_{j}}\Im\nabla_{x_{j}}\Psi(\tilde{X})|_{x_{i}\approx x_{j}}|}
\end{equation}
 where these are time averages over time scales typical of the local oscillation periods and
 spin labels have been suppressed.  The region of consideration around $x_{j}=x_{i}$ is the zone where incident flux is changing due to the scatterers.  For this energy scale, we are concerned with the region near the classical turning point of the two-body potential out to the region where the potential makes small changes for such kinetic motion.\footnote{This makes the Coulomb potential and any other potential where the range may extend larger than the interparticle separation a poor candidate for this model and further consideration in those cases is important.}  

If we think in terms of momentum flux balance this gives $\braket{\rho mv\cdot v}=2\mathcal{K}$ along each coordinate evaluated near the two body diagonals.  Here $v$ is the velocity of the currents $v=j/\rho$ along each one-body coordinate direction.  At higher energies we expect oppositely moving traveling waves to have no correlation and potential energy contributions to vanish so that half of the kinetic energy is in the form of such fluxes.  
This shows that we should have the same kinetic energy contribution for gases along each coordinate direction.  This condition does not depend on the particles having the same masses.

%For solids the 
%Thus, at an interface of two materials with different dispersion relations, we see a potential reason that the equilibration condition gives equal energies.    

A very simple picture of the equilibrated state we envision as a fine scale excitation over the ground state of the system where the oscillations' wavelength is much finer that the curvature of the potentials and ground state wavefunction.  Because of this, we expect it to ``fill in'' regions of lower potential energy function $\mathcal{U}(\tilde{X})$ much like water fills in a basin.  The kinetic energy density is
\begin{equation}
\label{eq1}
    \mathcal{K}(\tilde{X})=
    \begin{cases}
      E-\mathcal{U}(\tilde{X}), & \text{if}\ \mathcal{U}(\tilde{X})<E \\
      0, & \text{otherwise}
    \end{cases}
  \end{equation}
subject to the condition that 
\begin{align}
\label{eq2} 
\int dX^{N}[E-\mathcal{U}(\tilde{X})]\Theta(E-\mathcal{U}(\tilde{X}))=E_{thermal}
\end{align}
that implicitly defines $E$. This is reasonable as long as the vast majority of the the amplitude is not in the low KE density region where the function tails off.  For phonons we have a set of very anisotropic wells corresponding to orthogonal directions in $\mathbb{R}^{3N-5}$.  When we get excited occupancy in all directions the material tends to melt \cite{Reif}.  For realistic solids we expect that the lowest phonon modes will have high occupancy and this tails off to near zero occupancy for a finite fraction of states.  The thermal wavefunction we can expect to satisfy the high frequency equilibration condition in terms of the two-body diagonals of the normal coordinates: $u^{(i)}(\tilde{X})$ but not for the low occupancy states.  This means that this simple picture is only valuable for a solid when the vast majority of the thermal energy is in such high occupancy modes.  

%If the ground state energy density is $\mathcal{E}_{0}(\tilde{X})$, we expect the state of net thermal energy $E$ is given by the energy density
%
%\begin{equation}
%\label{eq1}
%    \mathcal{E}(\tilde{X})=
%    \begin{cases}
%      \epsilon(E), & \text{if}\ \epsilon(E)>\mathcal{E}_{0}(\tilde{X}) \\
%      \mathcal{E}_{0}(\tilde{X}), & \text{otherwise}
%    \end{cases}
%  \end{equation}
%subject to the condition that 
%\begin{align}
%\label{eq2} 
%\int dX^{N}(\mathcal{E}(\tilde{X})-\mathcal{E}_{0}(\tilde{X}))=E
%\end{align}
%that implicitly defines $\epsilon(E)$.  

We can not specialize our equilibration condition to describe situations where we expect the collection to have thermal meaning: 

{\bf Equilibrated Thermal State}: A wavefunction where there are a combination of both time varying currents and standing waves with typical wavelengths over a broad distribution that are much smaller than the scatterers and the mean free path, $\lambda_{typ}<<\sigma^{1/2},\lambda_{mfp}$, and the local oscillation distribution of the many body wavefunction is stable for long times over short time averages.\footnote{The case of electrons in a solid is slightly deceptive.  It seems that, at typical temperatures, these wavelengths are often rather long compared to the atomic, electron-electron separation or electron mean free path.  The decomposition of electron and core parts is artificial.  There is really only one wavefunction for the combination of them.  The typically short thermal wavelengths of the heavy cores determine the scale for both over time.  This distinguishes electrons in solids from fermionic gases where no such background of heavier objects exists.  The effects of this on thermalization will be discussed.}  %The energy density is modified as in Eqns.~\ref{eq1} and \ref{eq2} assuming smoothing is done on the scale of the typical wavelengths.  

%This suggests a kind of ballistic scattering reminiscent of ideal gases.  Degeneracy is a notion derived from noninteracting theory where products of single particle states give the general many body eigenstates.  The general many body analog still exists but as a less visualizable stiffness in the wavefunction against energy changes at low energies.  This condition or the appearance of long scattering lengths relative to the typical oscillations or relatively strong curvature in the potential, suggest that the thermal noise in our wavefunction may not be strong enough to occupy all the regions of sufficiently low potential with a density that is only a function of the local value of $V$.  

To continue, we consider that the kinds of matter we most associate with classical behavior are solids and gases.  (Liquids are still a subject of much debate.)  In the former case the oscillations are phonons.  In the latter, they are mostly unobstructed long range motions of individual particle waves.  These both give nearly free motions in terms of these respective bases.  Instead of giving an exact representation of the wavefunction on the true eigenbasis, we will utilize this property and describe it on the corresponding basis for this nearly-free approximation (NFA).\footnote{It is important to not use the true basis set for this description because we can start with an enormous distribution of energy eigenstates still attain a ``thermal state.''  The energy distribution in this basis will never change in isolation.  We use the NFA basis in this global fashion but are really only interested in it as a good local description of the wavefunction away from the two-body diagonals.  This is certainly not constant as the wavefunction evolves.  To the extent that the system allows a viable 3D description, the evolving expression on this basis describes the microscopic current oscillations to give a more physical approach to stochastic thermal motions.}  Specifically, for gases, we use a basis of free waves and for solids, we use a basis of phonons.  It is not clear if this is a completely general approach.  When excitations get large enough, as when temperature is large enough to cause appreciable expansion of the material and change in its elastic properties, the state is still built of many body eigenfunctions but they not decompose into products of the phonons as in the case of the low energy states.  Our motivation here is that, for such high frequency states, geometric scattering tends to locally dominate and so that such an NFA basis gives a good local picture of the wavefunction away from the two-body diagonal scattering regions.  It is such a basis that we presume always exists and will give meaning to the otherwise hard-to-justify ensemble averages and macroscopic thermodynamic quantities.  

%These are cases where there is a fairly accurate product decomposition of the (quasi) ground state and the nearby excited states (or an interesting range of excited states) in terms of long lasting quasiparticles.  For systems where this is not valid, presumably some similar effect occurs but it is not clear the best approximation of such a wavefunction even in the lowest energy state.  

As a specific case, let us attempt to encode the (free and spinless) MB gas into a trial thermal wavefunction.  Given the constant energy sphere in N-body momentum space we know that typical many body wavevectors $\tilde{k}=\{k_{1},k_{2},\ldots k_{N}\}\in\mathbb{R}^{3N}$ satisfy the MB distribution in its one body components.  We can construct a trial wavefunction
\begin{align}
\Psi=\hat{\mathcal{S}}\prod_{j=1}^{N}(e^{ i k_{j}\cdot x_{j}})
\end{align} 
Since this is an eigenstate, it has no temporal oscillations so is not suitable for our purposes.  Neither does it have the expected MB statistical distribuion.  

%An alternate encoding could be
%\begin{align}
%\Psi=\hat{\mathcal{S}}\prod_{j=1}^{N}\left(\frac{1}{\sqrt{N}}\left(\sum_{m=1}^{N}e^{ i k_{m}\cdot x_{j}}\right)\right)
%\end{align} 
%Nonsense
%\begin{align}
%\Psi=\prod_{j=1}^{N/2}(e^{ i k_{j}\cdot x_{j}})\prod_{j=N/2+1}^{N}(e^{- i k_{j}\cdot x_{j}})
%\end{align}
%Since this is an eigenstate, it has no temporal oscillations.  An alternate encoding could be
%\begin{align}
%\Psi=\prod_{j=1}^{N}\left(\frac{1}{\sqrt{N}}\left(\sum_{m=1}^{N/2}e^{ i k_{m}\cdot x_{j}}+\sum_{m=N/2+1}^{N}e^{- i k_{m}\cdot x_{j}}\right)\right)
%\end{align}
%GET THESE FACTORS AND INDICIES RIGHT!  ***  I THINK THE SYMMETRIZATION OP MAY CANCEL THIS.  IT IS BOSONIC.  
%This latter wavefunction has (desirable) oscillatory time dependence, since it is not an energy eigenstate, but also possesses phase correlation at $t=0$ that are undesirable.  

Using the one body kinetic result as motivation, let us consider a general Boltzman weighted sum on a free gas basis.  Let $\psi_{k}=e^{ik\cdot x}$ be the one body free states with energy $E_{k}=\frac{\hbar^{2}k^{2}}{2m}$.  (The volume normalization is chosen $\pi^{3}$ for simplicity).  
\begin{align}
\Psi= \hat{\mathcal{S}}\prod_{l=1}^{N}(\sum_{k\in \mathbb{Z}^{3}}\psi_{k}(x_{l})e^{-\beta E_{k}})=\sum_{\tilde{k}\in \mathbb{Z}^{3N}}e^{i\tilde{k}\cdot \tilde{X}}e^{-\beta \frac{\hbar^{2}}{2m}\tilde{k}^{2}}=\sum_{\tilde{k}\in \mathbb{Z}^{3N}}e^{(i\cdot \tilde{X}- \frac{\hbar^{2}\beta}{2m}\tilde{k})\cdot\tilde{k}}
\end{align}
This latter wavefunction has (desirable) oscillatory time dependence, since it is not an energy eigenstate, but also possesses phase correlation at $t=0$ that are undesirable.  
To eliminate the artificial phase correlations we choose a set of random phases $\theta(\tilde{k})$ and define our trial thermal wavefunction as 
\begin{align}
\label{psith}
\Psi_{th}= \hat{\mathcal{S}}\sum_{\tilde{k}\in \mathbb{Z}^{3N}}e^{(i\cdot \tilde{X}- \frac{\hbar^{2}\beta}{2m}\tilde{k})\cdot\tilde{k}}e^{i\theta(\tilde{k})}
\end{align}
where the phases are random up to the symmetry conditions implied by bosonic or fermionic symmetry.  (Spin labels have been suppressed in this discussion.)  %A similar result could be done for phonons in a solid but, due to the reorientation of equivalent oscillatory modes at each local maxima, the expression is more complicated.  

%In the case of gases, we have only included scattering as the driver of mixing that leads to a state like this in our NFA basis.  In the case of solids, we note that there are always nonlinearities, defects and external noise that prevent our SHO approximation of the oscillations as perfect.  Since the solid state is not even a true ground state, equilibration at any $\beta\ge 0$ is not truly possible.  
%
%*** Why do all states tend to head to this?  What is local meaning of NFA basis?   One body state approximation valid at higher energies locally.  Solids and phonons.  Some are high occupancy.  Some are empty.  Relate to quasiparticle typical state.  The equilibration requires the state be made of sufficiently many ess and have a large enough dE.  We throw out phase as in the Boltzmann equation?  

As evidence for equilibration among wavefunctions that tend to such a local description,  consider the \textit{interacting} dilute gas.  A distribution of high frequency random oscillations produce currents that interact with the hard core scattering centers in the manner of geometric optics.  In the two body CM frame, these centers gives the same dynamics as packets with energy and momentum conservation; the same dynamics that generate the classical MB distribution for hard spheres then apply here.  The big difference is in the delocalization of the many body wavefunction so that is cannot be thought of as a set of billiard balls on the microscopic scale.  Ultimately, even at such high frequencies, we know that there are eigenstates and superpositions that do not give us anything like MB or hydrodynamic behavior.  From a Green's function point of view, these are cases with constructive scatting interference.  The supposition here is that, at least for thermodynamics,  this is rather unusual and energy distributes itself rather uniformly over space and with a range of frequencies given by Eqn.~\ref{psith}.  We have presumed a unique local equilibrium distribution and this should be investigated further but, as in the classical case, expect it to be so.  

The conditions on the smallness of the wavelength versus the scatter sizes and separation extends this reasoning to the case of solids with phonons as discussed above.  In the case of liquids, the localized peaks presumably delocalize into a percolating structure that replace the lattice of well defined locations; some of the phonon modes being replaced by traveling modes that transport mass and allow greater penetration of vorticity.  The ``necks'' in the percolating structure could provide for scattering losses from asymmetry in the density buildup about them and the shape of these altered by electron bond strain.  This removes the need for aperiodicity to produced scattering which is important since we showed earlier that that notion of periodicity in a crystalline solid has no intrinsic meaning among transformations of the wavefunction corresponding to it \cite{Chafin}.\footnote{This is often overlooked because the electron part of the wavefunction for a solid with fixed cores is typically what is written down.  This function does have discrete translational symmetry.  However this part has only a limited role in determining the fluidity of a liquid.  }    %\ref{solids} 

%\section{Stability}
\subsection{Transitive Equilibrium}
So far we have not define the parameter $\beta$ which we anticipate will be related to the reciprocal of temperature.  The most basic feature of temperature is its transitive nature; equilibration of a body A with a body B and equilibration of B with C implies A is also in thermal equilibrium with C.  We can check to see if two such thermal wavefunctions with the same value of $\beta$ give a similar state.  
\begin{align}
\hat{\mathcal{S}}\Psi_{th}\Psi_{th}'&= \hat{\mathcal{S}}\sum_{\tilde{k}\in \mathbb{Z}^{3N}}e^{(i\cdot \tilde{X}- \frac{\hbar^{2}\beta}{2m}\tilde{k})\cdot\tilde{k}}e^{i\theta(\tilde{k})}\sum_{\tilde{k}'\in \mathbb{Z}^{3M}}e^{(i\cdot \tilde{X'}- \frac{\hbar^{2}\beta}{2m}\tilde{k}')\cdot\tilde{k}'}e^{i\theta(\tilde{k}')}\\
&=\sum_{\tilde{k}''\in \mathbb{Z}^{3(N+M)}}e^{(i\cdot \tilde{X}''- \frac{\hbar^{2}\beta}{2m}\tilde{k}'')\cdot\tilde{k}''}e^{i\theta(\tilde{k})}e^{i\theta(\tilde{k}')}
\end{align}
Where $\tilde{k}''=(\tilde{k}',\tilde{k})$ and similarly for $\tilde{X}''$.  
Assuming that the action of scattering mixes the phases sufficiently, this NFA basis representation of $\Psi_{th}$ gives an equilibration for two uniform density interposed gases.  From this we can define thermometry based on equilibrium with a particular chosen standard gas.

\subsection{Thermodynamics}

The case of homogenous state thermodynamics is of limited interest but really all that we are in a position to now discuss.  Inhomogenous stationary cases like heat transport and dynamic cases like sound waves and hydrodynamics generally require the validity of a 3D description so that we can apply the Navier-Stokes equations and chemical potentials.  When such a description holds and the extent to which it applies in the case of ultracold gases will be the subject of the next section.  

The homogeneous case gives a starting point to try to assign meaning to the ensemble averages and thermodynamic variables.  Pressure, temperature and entropy can all be defined in terms of the NFA basis.  For an interacting gas this gives the free gas pressure which is correct microscopically.  The two body interactions alter the net pressure on the edges of the trap by excluding regions from contact with the support of the wavefunction and replacing them with potential terms.  Entropy and temperature have similar microscopic meanings.  The larger scale values don't have an immediately obvious connection with the true density of states of the system.  For this reason we investigate these in terms of the true eigenstates of the system.

%**Give meaning to the thermodynamic quantities and give spatial variations in them that are product like.  Density variations lead to grand canonical like results.  

Our initial trial wavefunction of Eqn.~\ref{trial} was built on the NFA basis not using the true many body eigenstates.  The thermal equilibration condition leads to local configurations well described by this but obtaining the macroscopic averages requires us to do better.  We know that the NFA basis with the thermalization property led to a dominant contribution from a narrow distribution of states.  For similar reasons we assume the same is true on larger scales where the true basis gives a better description.  It is important to remember that the net wavefunction is in the same eigenstate distribution it started in.  The scales we are talking about are much larger than the mean free path but much smaller than the total system size.  

We extend Eqn.~\ref{trial} for the quasicontinuum to a typical state of the form
\begin{align}
\Psi_{th}=\int dE g(E) e^{-\beta E} e^{i\theta(E)}\Psi(E)e^{iEt/\hbar}
\end{align}
where $\Psi(E)$ is a suitably isotropic and random element of the constant $E$ sphere.  When the distribution is strongly peaked, the hyperarea, $\mathcal{A}(\Psi_{th})$, of the dominant energy surface is all we need to characterize the state for this purpose.  
In terms of this we can define the entropy as $S=k_{B}^{-1}\ln \mathcal{A}(\Psi_{th})$ temperature $\beta=(k_{B}T)^{-1}=\partial\ln \mathcal{A}/\partial{E}$ which lead to the usual thermodynamic relations and canonical averages without requiring the use of mixed states or any type of ``ensemble.''  Introducing additional constraints on angular momentum and other conserved quantities is important when computing results for macroscopic matter.  These are always well defined here although it is not immediately clear how to best do this and if history independent results must follow.  The most natural guess would be to restrict each of the $\Psi(E)$ to be a most probably choice with the restricted angular momentum.  This certainly deserves more consideration but is probably an involved topic on its own.

\section{Hydrodynamics}

\subsection{Convection and Vorticity in Solids}\label{convection}
In \cite{Chafin} we discussed the kinematic constraints of a solid and the nature of  microscopic (longitudinal phonon) excitations.  The equations of motion of a wavefunction are always linear but the motions of fluids and even acoustic waves in solids contains nonlinear advective terms.  It is illuminating to examine how this and the classical pressure and stress arise in the solid case briefly first before examining the fluid case.  We discussed the case of longitudinal oscillations in \cite{Chafin} but shear waves also exist and these require the presence and transport of vorticity.  The location and mobility of vorticity is an important topic in fluid dynamics but here we will argue it is also very important for a comparison of classical hydrodynamics with angular momentum and quantum systems ranging from ultracold bosonic and fermionic gases to superfluid Helium.  

We conventionally think of sound waves as particular linear combinations of phonons.  The weakness in this argument is that the core displacements in the phonon are assumed to be small changes in their equilibrium positions.  For larger solid with long wave sound of even modest amplitude, the displacements can be many atomic spacings.  There is no reason to doubt that such excitations can be built on eigenstates of the system but it is not clear that they can be made from one-body excitations in the quasiparticle limit of such modes with long lifetimes.  In this case, by ``lifetime'' we mean that duration of time that the solid can be well represented in the form 
\begin{equation}
\Psi (\tilde{X})^{(N)}\approx\hat{\mathcal{S}}^{f/b} \int d\tilde{X'}_{\perp}d{u} F(\tilde{X};\tilde{X}'_{\perp},u)\Psi(\tilde{X'}_{\perp})^{(N-1)} f(u)
\end{equation}
where $F$ is a kernel that includes two-body and higher corrections near the ground state and $\tilde{X}'_{\perp}$ gives the 3N-3 coordinate space perpendicular to $u$.  

The classical equation for solid motion is the Navier-Stokes (N-S) equations with strong local restoring stresses
\begin{align}
\label{NS} 
\rho\partial_{t}v+\rho v\cdot\nabla v&=-\nabla\cdot\Pi\\
\partial_{t}\rho&=-\nabla\cdot(\rho v)
\end{align}
where the internal stress tensor $\Pi$ provides energy and momentum conservation and provides restoring forces.  (Following convention we have used $\rho=nm$ here as the mass rather than the particle density.)  More accurate treatments include thermodynamic changes from local compression and their effects on temperature.  For gases this gives a large contribution to the speed of sound but for our discussion of solids we neglect it.  

As a first example consider the wavefunction of a solid, $\Psi_{class}$, consisting of N distinct atoms on a lattice as in \cite{Chafin} Eqn.~2 but without the symmetry constraints on the core locations.  This corresponds to a single localized peak in $\mathbb{R}^{3N}$ space corresponding to the lattice sites $\tilde{R}=\{R_{1}\ldots R_{N}\}$.  If we perform a displacement of the cores while preserving their localization we obtain a new set of lattice sites $\tilde{R'}=\{R_{1}'\ldots R_{N}'\}$.  We can map this into a 3D density function $\rho(x)=m\braket{R_{i}}$ where $\braket{R_{i}}$ is the volume averaged density of the cores on a scale much larger than their separation.  By using the local velocity of the cores and the bond energy density we can define $v$ and $\Pi$ to derive Eqn.~\ref{NS}.  This follows from the lagrangian form of the Schr\"{o}dinger equation since these are long lasting qualitiative states and in this limit we obtain the classical lagrangian.  
Since N-S are a nonlinear equations we expect that the corresponding evolution in terms of an eigenstate expansion must probe regimes of the many body eigenstate spectrum beyond any linear  quasiparticle picture.  It also implies that such a regime always exists in any system that gives classical N-S behavior.  

Now let us consider the simplest case of vorticity and solids: rotation.  A classical body undergoes rigid body rotation with velocity field $v=r\omega$ where $r$ is the distance from the axis of rotation.  This velocity field has the virtue of having the lowest kinetic energy of any velocity flow field with a fixed angular momentum.  Quantum systems with large amounts of angular momentum like superfluid Helium exhibit an Abrikosov vortex lattice \cite{Putterman} that give this as an averaged velocity field.  This suggests that we might expect a solid to have such a hidden lattice of rotation somehow hidden in its corresponding many body wavefunction $\Psi_{class}$ however we will see that this can be very far from the truth.  We saw in \cite{Chafin} that being specific about the kinds of wavefunctions that specify classical matter removed the apparent problem with macroscopic superpositions by giving low energy long lasting partitions of the observers evolution and that naive superpositions created huge energy barriers that lead to unphysical states.  Here we will see that the way vorticity can enter the wavefunction of such a classical body generates very little energy change unlike the usual quantum fluid examples we are familiar with.  

Let us continue with our example of distinct particles in the solid as above.  The state of this system is represented by a single peak at $\tilde{R}$ that now evolves according to the classical orbits $\dot{R}_{i}(t)={\omega}\times R_{i}(t)$.  This generates a 1D loop in $\mathbb{R}^{3N}$.  This loop induces a 2D surface but, in such a high dimensional space, it  has no unique normal to orient a vortex line about the center.  We can, nevertheless, choose any one of many vortex lines through the middle it with a (typically huge) winding number that generates the orbit in a period about the loop of $T=2\pi/\omega$.  The vortex cuts through the
low amplitude tail of the wavefunction so has very little energy contribution to the system.  Furthemore, there are many such vortices that can generate the same behavior.  In fact, many vortices could be used as long as they generate the uniform orbital motion around the 2D loop.  If we now allow the atoms that make up the body to be identical (or just as problematically, allow them to be composed of the same type of constituent fundamental baryons and leptons) we now symmetrize and generate a set of $N!$ peaks and the same multiplicity of new vortex lines.  In a solid there are many low amplitude regions between the cores so it is energetically favorable for them to fish their way through these percolating ``interstitial'' gaps.  

Now consider a cubic solid that has undergone a shear deformation and is released at $t=0$.  The net angular momentum of the solid is zero but the motion is now a combination of shear and rotation so that relative motions induce opposing vorticity that is created and destroyed over time.  We can still use the above method to track the core locations and make statements about the kinds of vorticity distributions allowable.  In particular, the constant density of the material implies that vorticity only passes in and out of the surface rather than being created.  This means that we can expect apparently singular domains of vorticity at the edges of the body in the low amplitude tails of the wavefunction.  

Even though it seems like there is no net vorticity, the presence of opposing vorticity have dynamical effects are are important especially in classical fluid dynamics.  The enstrophy is defined as the measure of the square of the voriticity.  For a constant density fluid, this has kinetic energy contributions.  However, for our solid, it seems that the low amplitude regions of the wavefunction induced by the strong binding forces that induce relative localization and motion constraints among the cores are so much more important than any vorticity contribution to the kinetic energy that vorticity is simply created, destroyed and moved about as needed with little effect on the classical motions.  Gases will not have such features and the role of vorticity here will be different.

\subsection{Classical Gases}
The distinction between classical gases and familiar quantum gas cases is central to the upcoming discussion on thermalization and the validity of hydrodynamics for ultracold gases.  For this reason we will enumerate some of the differences in the observed behaviors and what corresponding differences the many body wavefunction describing each must exhibit.

Consider an isolated 1~kg solid.  The delocalization of an atom versus that of the CM is related by $\sqrt{N}\delta X_{a}=\delta X_{CM}$.  Assuming each atom is localized to $\delta X_{a}\approx10^{-11}$m we have the CM localized to $\delta x_{cm}\approx 10$~m.  The delocalization rate of the CM is therefore $v_{del}\sim 10^{-33}$~m/s.\footnote{It is surprising that the CM is not extremely well localized based on this assumption but the kinematics of large bodies can be unaffected by it.  Placing two such bodies adjacent to each other encounters no overlap or restrictions from it since the interactions come from the lack of localization in each individual coordinate label.}
In contrast, the delocalization rate of a lone atom with the same initial localization is $v_{del}\sim10^{4}$~m/s.  Thus, while the atoms of a solid maintain the picture of being classical very well, the gas leaves the category of ``classical'' rapidly and expands into a very delocalized object where strong correlations are necessary to avoid the large overlap energies we obtained earlier from ``naive superpositions'' \cite{Chafin}.  The kinetic oscillations will still dominate the gas as they are much more energetic than the curvature of delocalization.  In one sense, this is heartening, if the gas were to stay very localized then it would cast doubt on the relevance of the entropy values derived from the Sakur-Tetrode equation.  In another sense, it is distressing because it makes it hard to justify the 3D parcel assumptions we generally use to build thermodynamics and hydrodynamics.  Being so energetic in the relative coordinate directions suggests that there is very little energy cost to generating vortices that enter the support of the wavefunction.  

We can make a first estimate this by using the thermal wavelength to bound the size of norm exclusion about a vortex core.  In the low temperature bose gas case the correlation length specifies the damping of the order parameter about a vortex \cite{Putterman}.  Here the vorticity does not experience the same tendency to correlate over all particles but for our estimate here we want the density to truly get driven to zero when we symmetrize so we assume an N-fold vortex product.  Such a ``vortex line'' of length $d$ generates a volume exclusion of $\sim  \lambda_{th}^{2}d$.  The local energy cost for this is very little kinetically since the local oscillation curvature of the attenuation about the vortex is comparable to the thermal curvature it replaces.  The density change must be compensated for in the overall compression of the gas so, assuming each vortex\footnote{We will reexamine this assumption more carefully in Sec.\ \ref{angular}.} gives angular momentum $N\hbar$,  we have a net hidden quantum contribution to the thermodynamic energy of $\Delta E\approx P ( \lambda_{th}^{2}d)(L/N\hbar)$ where $d$ is a measure of the extent of the cloud and $L$ is its net angular momentum.  From this we conclude that $\Delta E\ll E=\frac{3}{2}k_{B}T$ when the rotational period of the cloud, $\Omega$, satisfies $\frac{k_{B}T}{\hbar}\gg \Omega$.  (If we have residual quantum enstrophy this gives another increase in the energy correction.)  For a room temperature object this gives $10^{12}\gg\Omega$.  It is evident that only very small and tightly bound objects can ever obtain such a rate of rotation.  When $\Omega\sim k_{B}T$, which occurs for $T\sim10^{-8}$ for KHz frequencies, the quantum correction to the thermal energy of the system becomes comparable to classical thermal one which is when visible coherent vortices dominate the motion instead of rigid body rotation.  In both cases the angular momentum is due to vortices in the cloud but in the high energy case there is no favored coherence of them so that angular momentum must increase in $\sim N\hbar$ steps rather than just $\hbar$ steps.  

In principle we will see there is no quantum limits on the actual size of vorticity due to possibilities for superpositions but there is always a ``vorticity penetration cost'' that dominates in the lower energy states where the ground state curvature is large relative to excitations.  How vorticity appears in higher energy wavefunctions to give classical hydrodynamics is the central topic of the next sections.

The effect of collisions of gas atoms with solid walls is known to be profound.  It was observed in the 1880's that the viscous flow of gases in narrow tubes can only be explained if there is a nonspecular reflection of gas particles at a solid surface \cite{Jeans}.  The results are consistent with adsorption and reemission of gas particles as thermalized with the surface.  A delocalized gas molecule undergoing an effective measurement by the surface would effectively be part of the phonon structure of the solid at the end of it.  Ejection follows from the state of the surface being unable to bind the molecule even at rest rather that from a conservation of momentum at the surface.  A highly correlated history of collisions induces some constraint on its history in that slice.  Whether this is a strong enough effect to induce a well defined 3D hydrodynamic gas flow in each slice is an important question and a possible future test for this model.  

In the absence of such solid material, the gas will continue expanding to even much larger distances to create a wavefunction that has little in common with the classical billiard ball picture of a gas.  The distinction is so strong that it is hard to see how calculations based on this model can be considered to make any statement about the system.  It turns out that things are not that bad but only after a more serious consideration of the system based on its actual state as a wavefunction.  Thermal properties seem to agree, however, the resurrection of classical hydrodynamics is rather complicated.  We follow our example with solids and write down the simplest plausible many body wavefunction consistent with classical hydrodynamics and discuss some conditions under which it would evolve according to the hydrodynamic limit.  

\subsection{Quantum Gases}\label{quantumgas}
The experimental quantum fluids are superfluid $^{4}$He, $^{3}$He and ultracold gases.  There are other examples of experimental quantum systems like superconducting electrons and exciton-polariton condensates that can exhibit migration but neither exhibit the clear  fluidity of hydrodynamic systems.  The telltale features of superfluid behavior are irrotational motion, specifically the presence of vortices, a lack of viscosity and two-fluid behavior.  In the case of Helium, the two-fluid model has been well verified \cite{Putterman}.  Damping has been measured in particular configurations like the oscillating plate Andronikashvilli experiment and the case of vibrating wires.  It is often stated that these experiments measure viscosity.  Damping can be calculated by the phonon and roton scattering model of Landau and Khalatinikov \cite{Khalatnikov:65} however viscosity is more than just a measure of damping.  It is defined by the N-S equations and places very particular constraints on how vorticity can move and enter the fluid.  In the constant density fluid, the vorticity transport theorem \cite{Batchelor} implies vorticity is advected and can only enter or leave through the boundaries.  In Sec.\ \ref{ultracold} we will examine the possibilities than nonequilibrium history dependent behavior is persisting and what small effects might be evident in the cloud shape from them including a tendency to keep vorticity out of the higher density regions.  

The two-fluid model describes a fluid as having a normal and superfluid component where the superfluid part has no viscosity.  As $T\rightarrow0$ the normal component vanishes yet these damping processes persist and the critical velocity in small tubes remains finite.  The damping of flow in larger tubes and flow rates what phenomenologically explained by Landau due to quasiparticle excitations.  This explanation is largely no longer believed as pictures of superfluid turbulence have become more clear.  Thus the case of superfluid Helium seems to still have some mysteries despite the successes of the two-fluid model and there is no fundamental understanding of how the two-fluid model arises.  

These considerations are important because ultracold bosonic gases have exhibited long lasting vortices.  This suggests superfluidity and that a two-fluid model might apply to them however, these vortices often decay leaving one to wonder what is happening to the angular momentum they possess.  The possible existence of quantum limit on viscosity \cite{Kovtun:2004de} is a pressing problem and has attracted much theoretical and experimental attention with the study of these gases playing a central role.  For this to be a valid approach, thermodynamics and hydrodynamics should provide a well defined description of them.  

The most simple and successful treatement trongly repulsive bosons at low energy is the Gross-Pitaevskii (GP).  The simplest treatment of GP is to assume that the many body wavefunction is a simple product $\Psi=\prod_{i}^{N}\psi(x_{i})$.  Variational methods then give the evolution equation
\begin{align}
i\hbar\partial_{t}\psi(x,t)=\left(-\frac{\hbar^{2}}{2m}\nabla^{2}+V(r)+\frac{4\pi\hbar^{2}a_{s}}{m}|\psi(x,t)|^{2}       \right)\psi(x,t)
\end{align}
where the last term which could be summarized as the interaction strength ``g'' is written in terms of the two-body scattering length $a_{s}$ and the normalization has been chosen $\int |\psi|^{2}dx=N$.  With this normalization $\psi$ is usually called the order parameter and labeled $\Phi_{GP}$.  It is distressing to have a nonlinear equation appear but the price of forcing a product function solution.  A more realistic function would have the form
\begin{align}\label{product}
\Psi(\tilde{X'})=\int d\tilde{X} F(\tilde{X'},\tilde{X})\prod_{i}^{N}\psi(x_{i})
\end{align}
The interaction strength is really a measure of the interaction contribution at the near contact potential and the kinetic energy induced by the curvature as summed up in the kernel $F(\tilde{X'},\tilde{X})$.  This is important since the true interaction between particles is often attractive.  We can obtain effectively repulsive effects by having the system in an excited branch of the two body potential's spectrum.  

Being in a low energy state suggests that the wavefunction is strongly limited in the motion it can exhibit.  If we have a cloud in a spherical harmonic trap and make small quasi-static deformations of it through a one-parameter set of quadrupolar deformations, we can obtain a set of wavefunctions $\Psi(\tilde{X};\alpha)$ where $\alpha$ gives the deformation.  A deformation that is then released in the spherical trap can then evolve over this set to a good approximation so long as $E_{j}\ll E_{s}$ for all time.  This gives a solution $\Psi(\tilde{X};\alpha(t))$.  The many body $\Psi$ can be described in terms of a limited set of variables like the one-body density $\rho(x)$, a one-coordinate projection $\psi(x)=\psi(x_{1})=\Psi((\tilde{X})|_{x_{2}=c_{2}\ldots x_{N}=c_{N}})$, the best fit product expansion as in Eqn.\ \ref{product}, or even the single parameter $\alpha$.  Any equation of motion in terms of these variables must eventually fail since the $\Psi$ has current terms and so is not going to stay well describable by such a limited class of function forever.  Since the Schr\"{o}dinger equation can be written as a hydrodynamic equation with a quantum pressure term it is not surprising that we can get a linearized Euler equation for the dynamics in terms of $\psi(x)$ induced $\rho(x), v(x)$ valid for some finite time.  This is expressed in Fig.\ \ref{fig: NS2}.  The extent to which such a situation holds for more general gases  the subject of the following sections.  
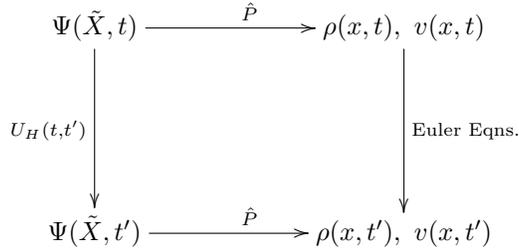
\begin{figure}
\[
\xymatrix @C=5pc @R=5pc{
\Psi(\tilde{X},t) \ar[d]_{U_{H}(t,t')} \ar[r]^{\hat{P}} & \rho(x,t),~v(x, t) \ar[d]^{\text{Euler Eqns.}} \\
\Psi(\tilde{X},t') \ar[r]^{\hat{P}} & \rho(x,t'),~v(x, t') }
\]
\caption{For a near ground state of an interacting Bose gas the kinds of adiabatic deformations and release in a trap and the stiffness of such a wavefunction implies near commutativity of the diagram.  In contrast, a noninteracting gas makes this exact since the induced many-body currents cannot transfer energy to its transverse directions.  }\label{fig: NS2}
\end{figure}

Another approach to such abbreviated ``order parameters'' to describe $\Psi$ is to look at it along a particular subslice.  For bosons we could use the value of $\Psi$ near the many body diagonal $x_{1}=x_{2}=\ldots x_{N}=x$.  The function must near vanish there due to interactions but we can look at an $\epsilon$ displacement from it $\psi_{d}(x)=\Psi(x+\epsilon,\ldots x+\epsilon)$.  This is also possible for a fermionic wavefunction as long as we don't evaluate it on a node $\psi_{d}^{(f)}(x)=\Psi(x,x+\epsilon_{1},x+\epsilon_{2},\ldots x+\epsilon_{N})$ where $\epsilon_{i}\ne\epsilon_{j}$ for all $i\ne j$.  This can be made unique by choosing the many body directions that give the largest local amplitude.  
Of course, such an approach is most interesting if we can give an equation of motion valid for reasonably long times in terms of it or relate it to some measurable quantity.  For large deformations or time changing potentials or interactions, it will often be the case that such a description is not meaningful since the evolution of $\Psi$ will access many more degrees of freedom on much shorter time scales.  

In the favor of such an approach for bosons is that the GP equation for ultracold bosonic gases gives excellent static description of cloud densities.  
Dynamically, it has given qualitative and sometimes quantitative descriptions of such clouds \cite{Dolfovo}.  In the case of small oscillations of boson and fermion clouds near unitarity, the hydrodynamic approximations for static cloud shapes works  very well and predicts the small oscillation periods quite accurately at least for the lowest modes \cite{Stringari:2004}.  This is rather profound.  A classical gas as a set of billiard balls defines mean free path, collision times, cross sections and the like in a very intuitive manner.  The quantum case, as we have seen, allows rapid delocalization of  localize packetsd into a cloud where correlation effects can dominate.  Even in the high temperature regime it is not clear that we should get Navier-Stokes evolution and that the huge cardinality of degrees of freedom in a general many body wavefunction should condense to the very limited set corresponding to a cloud with well defined temperature, velocity, density etc.\ that are purely 3D variables.  Microscopically we expect that this is a single wavefunction with a well defined phase moving with only singular sources of vorticity.  We know a wavefunction can exhibit perfect stationary or periodic motions at any energy by producing linear combinations of eigenstates, yet hydro and thermodynamics give a vastly restricted class of motions.  To date there has been no derivation of classical hydrodynamics from a wavefunction based approach.  It is not even obvious how to choose appropriate initial data for such a description.  

Shortly, we will give a proposed explanation for N-S behavior as a suitable limit for high temperature gases and a non-hydrodynamic explanation for why and when the linearized N-S equations should give a valid description for oscillatory modes in bosonic gases.  Damping of these modes provides an additional challenge usually tackled through linear response theory and the Kubo formula.  These can be thought of as scattering based approaches.  For a fixed external potential and interaction strength, the system could also be considered as expanded on a basis of eigenfunctions.  This should give a consistency check on linear response theory since the end results must be consistent.  The Kubo formula for dissipative response has been very successful yet there are longstanding serious doubts about the validity of its derivation \cite{Kampen}.  No conclusive resolution of these problems have been obtained.  Linear response theory is generally applied to get response functions and transport coefficients.  In the case of hydrodynamics, a N-S model with possible gradient expansion is assumed and the coefficients are derived.  Classically, higher order expansions tend to have convergence problems \cite{Cohen}.  The Kubo approach generates fractional order expansion coefficients that cannot be mapped onto any gradient expansion.  Fractional derivative equation might hold some promise for resolving this but tend to introduce many unfavorable behaviors on their own.  Generally, this situation is described as the case when hydro ``breaks.''  However, the aspect of this problem that is generally ignored is when the N-S expansion is even valid at lower order for such gases.  For it to be so, important correlation properties must hold and persist.  This would then be a specific model to give an eigenstate based consistency check on the linear response approach.  

\subsection{Angular Momentum and Vorticity}\label{angular}
In quantum mechanics we learn that angular momentum is quantized in units of $\hbar$.  This is evident from eigenstates of the Hydrogen atom.  Photons don't have a wavefunction and classical electromagnetic waves exhibit a kind of ``angular momentum paradox'' due to the fact that classical spiraling wave solutions seem to posses zero angular momentum.  However, for photons, we use second quantized solutions with angular momentum given by the helicity of the wave derived from the operator rules \cite{Schweber}.  In the case of superfluids we append the superfluid phase rule $v_{s}=-\frac{\hbar}{m}\nabla \phi$ to the two fluid model and verify integral increases in angular momentum corresponding to all particles in the superfluid fraction contributing $\hbar$.  

For a general wavefunction, is it clear that angular momentum requires a current hence an advancing phase (so $E_{j}\ne0$).  Topologically, this implies there are 3N-1~D hyperlines where the phase evolves about with diverging velocity near them, the amplitude of the wavefunction must vanish there.  Thus, angular momentum necessitates the presence of vorticity.

Angular momentum is always determined about a fixed base point.  
In the study of rigid bodies, the parallel-axis theorem tells us we can compute the angular momentum of a system about their CM and then the bulk motion of these bodies about the base point.  For a general wavefunction let us consider the implications of shifting the base point on the angular momentum of the system.

For an Abrikosov vortex lattice in strongly interacting bosonic systems, the collection of vortices rotate like a rigid body and the velocity advances, when averaged over scales larger than the coherence length, to that of rigid body rotation.  Consider a wavefunction with uniform support in a cylinder of radius $R$, height $H$ and attenuation at $r_{0}\approx0$.  The density is $\rho=(\pi H(R^{2}-r_{0}^{2}))^{-1}$.  
If we have a single vortex line down the central axis the velocity field is $v(r)=v_{s}\frac{R}{r}$ where $v_{s}$ is the velocity at the surface then the angular momentum is $L=\int dm(r) v(r) r=\int (m \rho H ~2\pi r dr)(v_{s})r$.  Since this field corresponds to a wavefunction $v_{s}=n\frac{\hbar}{m r}$ where $n$ is the winding number so we have $L=n\hbar$.  The kinetic energy is $K=n^{2}\frac{\hbar^{2}}{m(R^{2}-r_{0}^{2})}\ln(\frac{R}{r_{0}})$ which tends to the classical result for a ring as $r_{0}\rightarrow R$.  The logarithmic trend vanishes in the limit of many closely packed vortices but for dilute vortex fractions such considerations are important.  Thermodynamic quantities are generally extensive in the size of the sample.  Angular momentum is not but the ratio of it and kinetic energy for a solid are nicely related by $K/L=\frac{\Omega}{2}$.  For our irrotational rotating cylinder the ratio is $K/L\approx\frac{n\hbar}{mR^{2}}\ln(\frac{R}{r_{0}})$.   

If we displace the vortex from the center of our cylinder (moving the attenuation radius with it) so that the phase pattern remains uniform, both the angular momentum and kinetic energy are altered.  To measure the angular momentum we must make a choice between still using the center of the cylinder and using the new center of the vortex.  If we keep the base point at the center of our distribution and displace the vortex radially then the angular momentum of the system drops to zero as the vortex moves to infinity.  If we, alternately, keep our base point at the vortex center then the angular momentum decreases similarly.  As the vortex leaves the support of the density and approaches large distances, the cloud picks up a net motion about the vortex center.  In neither case is angular momentum (nor linear momentum) conserved.  Since the vortices are topological objects that can combine with vortices of opposite helicity but, otherwise must enter an leave the support intact, changes in the angular momentum must occur by changes in their curvature and orientation, the local phase gradients that wrap around them or the density of the norm.  Naive shifts in vortex structures are generally forbidden by angular momentum conservation.  

Regardless of what what base point we choose, the kinetic energy is, of course, the same which reminds us that the ratio of $K/L$ generally won't have any coordinate invariant meaning.  The angular momentum per vortex is now no longer $\hbar$.  How can this be?  The state is clearly not an eigenstate since there is now a transverse velocity field at the boundary.  The angular momentum was never really a localized quantity.  We only associated it to the vortex since it has an obvious and precise location.  The phase fronts do not need to be uniform as in the cylindrically symmetric case.  General superpositions are allowed so that the many body vortices can give $L<N\hbar$. 
These all make evident that intuition we have gleaned from superfluids or symmetrical quantum eigenstates are prone to lead us astray.  The interesting questions are how classical and superfluid motion arise from such a general wavefunction.  
It is now clear that we must now be more careful in assuming that there is a simple relation of the form $L=N_{vor}\hbar$ where $N_{vor}$ it the number of vortices in the sample.  If a uniform lattice exists with vortex separation distance given by $d$, there is a smoothed flow that gives rigid body rotation as is evident by computing the circulation about symmetrically nested circles.  

We are now in a position to compute deviations from the rigid body result.  For a uniform density rotating disk $L=\int d\theta \int \rho~r~dr (v r)$ where $\rho=M/\pi R^{2}$ and $v=\Omega r$ so $L_{class}=\frac{1}{2}M\Omega R^{2}$.  If we consider our vortices are made of nested rings at radii $r_{k}$ spaced by $d$, the circulation in the integral is altered since $v r=\Omega r_{k}^{2}$ for $dk=r_{k}<r<r_{k+1}=d(k+1)$.  The resulting angular momentum is 
\begin{align}
L=\int (r~dr) d\theta \rho~ (v ~r)%=\int dr d\theta \frac{M}{r} (\Omega r_{i}^{2})
\end{align}
where the approximation $v=\Omega r_{i}\frac{r_{i}}{r}$ between every pair $(r_{i},r_{i+1})$ gives the \textit{exact} circulation along every closed loop in this interval.  The resulting integral becomes
\begin{align}
L&=2\frac{M\Omega}{R^{2}} d^{4}\sum_{j}^{s} j^{3}\\
&\approx \frac{1}{2}M\Omega^{2}d^{2}\left( s^{2}-2s \right)
\end{align}
where $s=R/d$.  
Neglecting density changes from attenuation near the vortices, the classical result is excessive by $\Delta L\approx 2\frac{d}{R}L_{class}$.  Similar corrections exist for kinetic energy that is actually in excess of the classical result.  Notice that this correction is only small because of the large number of vortices.  A few widely separated vortices give energies that are not even independent of the size of the system.  

The case of interacting (repulsive) bosonic gases are the most dramatic in terms of angular momentum because these gases can exhibit visible vortices and vortex lattices like we see in superfluid Helium.  This is the opposite limit of a high temperature gas since the thermal oscillations are now small compared to the ground state curvature.  This leads us to expect the kinetic and potential energy cost of introducing vorticity will be much larger hence give different behavior.  
In the noninteracting case, we can have a product of individually evolving one-particle functions that each have arbitrary vortex structures.  The role of interactions is to force the many body wavefunction to cause these structures to correlate so that these give a visible density drop in the one body density function $\rho(x)$.  Using one-particle language we would say that every particle shares in the same vorticity structure.  If the vortex is at the center, this corresponds to $L=N\hbar$ where $N$ is the total number of particles.  
If we model two body interactions with a set of Jastrow-like corrections the function looks like
\begin{align}\label{eqn:vortex}
\Psi(\tilde{X'})=\int d\tilde{X} F(\tilde{X'},\tilde{X})\prod_{i}^{N}\psi(x_{i})
\end{align}
where $\psi(x)$ encodes the phase and vorticity structure and $F(\tilde{X'},\tilde{X})$ gives the curvature due to the interactions.  

The repulsive interaction is greatest at the two-body diagonals.  If the vortex structure was built from a set of functions that did not have vorticity at all the same points we could still symmetrize it but the cores would not be falling on these repulsive diagonals.  If we used only $n<N$ of the particles in the system then a fraction of the $~N^{2}$ diagonals would not intersect with these diagonals.  Even if this was only $\sim N^{-1}$ particles there are still $\sim N$ diagonals with amplitude not vanishing except for the damping action of the kernel $F$.  We have implicitly considered each vortex to have winding number $n=1$ (often the most stable case) but, in this ``one-body'' picture of the many body vortex, we also have to consider the ``occupancy'' of these vortices.  Implicit in the two-fluid model is the the superfluid component has ``full'' occupancy and, for ultracold bose gases, the stable states do as well.  This observation tells us that it is more accurate, for superfluids, to describe the vortices of having full occupancy rather than quantized angular momentum since we saw that sifting these vortices relative to the fluid surface changes it when we leave the base point fixed.\footnote{We will soon see that high temperature gases tend to a coarse grained irrotational product function to reduce scattering.  Vorticity is less expensive for these gases and the higher velocities are short range so not necessarily so constrained.  For strongly interacting gases, similar scattering effects may explain why vorticity does get so restricted.}  
A better understanding of this might lead to a non-phenomenological description of the two-fluid model in terms of the specific dynamics of a many body wavefunction.  

As we noted above, displacing the vortex from the center of an axially symmetric cloud, reduces the angular momentum of the system.  The coherence effect seems to be more a statement of the orientation of the many body vortices due to the interactions than one of angular momentum quantization itself.  This begs the question of what happens as the cloud relaxes as this vortex shifts or if less than such an optimal amount of angular momentum is initially imparted.  

If the angular momentum is low enough of a cloud in an axially symmetric trap, we can get surface waves which correspond to vorticity at infinity or at finite distances in the low amplitude tails of the distribution.  These vortices don't suffer the same vortex penetration costs.  Such surface waves are familiar in both bosonic and fermionic clouds.  What is unclear is what happens to it as these oscillations seem to settle down.  The angular momentum must be preserved.  In a classical gas, we expect the final state to give rigid body rotation as predicted by viscous N-S evolution.  In the case of interacting bosons, this seems to require vortices to penetrate the whole cloud uniformly like a vortex lattice or to be ``uncorrelated'' in the sense above.  This gives large kinetic energy contributions without the corresponding decrease in interaction energy.  An energetiically more favorable situation would by to allow some angular decorrelation in the picture of one-body states in Eqn.\ \ref{eqn:vortex}.  If we use the one-body functions $\psi(x_{i};\theta_{i}(t))$ as elliptically deformed rotating states where $\theta_{i}(t)$ indicates the angle of the major axis of the $i$th function then a final state such wavefunction is
\begin{align}\label{eqn:rot}
\Psi(\tilde{X'})=\hat{\mathcal{S}}\int d\tilde{X} F^{\star}(\tilde{X'},\tilde{X})\prod_{i}^{N}\psi(x_{i};\theta_{i}(t))
\end{align}
The choice of $\theta_{k}(t)=k\frac{2\pi}{N}+\Omega t$ gives an axially symmetric one-body density function for $\Psi$ that retains the angular momentum of the state with $\theta_{k}(t)=\Omega t$ for all $k$.  ($F^{\star}$ indicates that the kernel we use to include correlations may be somewhat different than the one used for the true ground state.)  
There is a reduction in the interaction energy at the surface due to reduced interaction energy.  This provides a distinct picture than the GP equation for angular momentum to exist and ultimately enter the bulk of the cloud in more observable form.  Since the overlap of the one particle wavefunctions gives the interaction energy.  The decorrelation of these semi-major axes reduces the repulsive interaction so that the cloud radius will contract compared with the simple rotation of the one body functions.

Due to interaction effects and the possibility of hidden phase gradients it would be convenient to have a measure of the number of one body wavefunctions that overlap.  
We can quantify the ``number density'' in this model 
in a fashion other than the usual GP order parameter and independent of the one-body density function $n(x)$.  Let 
\begin{align}
\mathcal{N}(x)=\frac{\sum_{i}^{N}|\psi_{i}(x)|^{2}}{|\psi_{max}(x)|^{2}}
\end{align}
where $\psi_{max}(x)$ specifies the index in the sum of Eqn.\ \ref{eqn:rot} that gives the largest norm at $x$.  An alternate definition that is model and basis independent is 
\begin{align}
\tilde{\mathcal{N}}(x)=\begin{cases}N\cdot f_{N}\left(\frac{\int d\tilde{X} {|\Psi(x_{1}\ldots x_{k}\ldots x_{N})|}\sqrt{\rho(x_{1})\rho(x_{2})\ldots\rho(x_{N})}\delta(x-x_{1})}{\rho(x)}\right)
 &\mbox{if}~ \rho(x)\ne0\\
0&\mbox{if}~\rho(x)=0\end{cases}
\end{align}
 $f_{N}(y)$ is a function that sends $y=1\rightarrow1$ so that product functions give $\tilde{\mathcal{N}}(x)=N$ for values of $x$ where $\rho\ne0$ and $f\rightarrow\frac{1}{N}$ for functions corresponding to well defined locations at different points.  For example, if we used a function $|\Psi|^{2}=\hat{\mathcal{S}}\sum_{j}^{N}\delta(x_{j}-d\,j)$ then $\mathcal{N}(x)=1$ at $x=d\,j$ for all integers $0<j\le N$ and zero elsewhere.  

In the case of a product function, of identical functions $\mathcal{N}(x)=N$ everywhere.   
These functions measure the the extent of spatial correlations.  Greater correlations allow hidden irrotational motion that can hide angular momentum from one-body functions like $\rho(x)$ or the GP order parameter $\Phi_{GP}(x)$.  If $\mathcal{N}(x)=1$ everywhere the GP solution is expected to be accurate.  A local value of $\mathcal{N}(x)<1$ indicates that a distribution of phase gradients can correspond to that location so gives a kind of bifurcation of the order parameter allowing an apparent rotational component to the velocity field driving fluxes of $\rho(x)$.   This suggests an alternate approach to higher corrections to the GP model but we won't pursue this further here.  Ultimately, we would like a measure that tells us when the vortex penetration cost is favored versus leaving angular momentum in surface oscillations about the surface.

\subsection{Hydrodynamic Kinematics}

Just as we worked to establish a set of plausible initial data for the wavefunction of a solid classical body that was consistent with the usual phonon and solid state electron orbital calculations, we would like to give some plausible subclass of wavefunctions that correspond to a gas that is thermalized and hydrodynamic.  In the case of the Gross-Pitaevskii equation for a condensed bosonic gas we often consider the order to parameter to be a scaled copy of a single particle wavefunction and all the particles to be ``in the same state'' as in Sec.\ \ref{quantumgas}.  This is a manifestly single body descriptions of the situation.  It is not manifestly clear this makes sense.  After all the ``g'' that describes the interaction strength (usually in terms of the scattering length) typically is a subtle combination of both potential energy of the interactions and kinetic energy of the wavefunction oscillations about the two-body diagonals.  One can derive a static version of the GP equation from the result of LHY \cite{LHY:57} or Bogoliubov \cite{Bogo:57} theory by applying mean field theory.  Extensions of this to time dependence, especially with higher order corrections, as in the case of Bogoliubov-Hartree-Fock theory, introduce unphysical gaps in the energy spectrum or violate conservation laws \cite{Yukalov}.  Despite these efforts, experiment seems to show that the case of very low T bosonic systems are very correlated systems, rigid enough that they only require a single order parameter for their description.  Persistent vortices can occur and hence these gases are often dubbed  superfluid, often leading to the inference that the two fluid model is applicable.   

What about the high temperature limit?  What enforces such a limiting rigidity on the system so that hydrodynamic variables are relevant.  In the case of contact potentials, Werner has come to the conclusion that most eigenstates at high energy are noninteracting \cite{Werner:2009}. This is of little help for seeking how generally and fast a hydrodynamic state is established.  In the noninteracting case we can assign every atom a wavefunction corresponding to a different flow.  Furthermore, we can superimpose these.  The questions we can ask are: Do most such states tend to ``equilibrate'' to a hydrodynamic state?  Does only a special subset of wavefunctions correspond to the kinds of flows we observe, perhaps by starting from equilibrium and then initially driven by classical 3D objects and forces with their associated kinematic constraints?  As in the case of thermodynamics, must any gas wavefunction tend to a recognizable hydrodynamic state on its way towards long term equilibration?  

Specifically we can ask when the diagram in  Fig.\ \ref{fig: NS} commutes.
\begin{figure}
\[
\xymatrix @C=5pc @R=5pc{
\Psi(\tilde{X},t) \ar[d]_{U_{H}(t,t')} \ar[r]^{\!\!\!\!\!\!\!\!\!\!\!\!\!\!\!\!\!\!\!\!\hat{P}} & \rho(x,t),~v(x, t),~ T(x,t) \ar[d]^{\text{Navier-Stokes}} \\
\Psi(\tilde{X},t') \ar[r]^{\!\!\!\!\!\!\!\!\!\!\!\!\!\!\!\!\!\!\!\hat{P}} & \rho(x,t'),~v(x, t'),~ T(x,t') }
\]
\caption{Mapping of a general many body wavefunction onto classical hydrodynamic variables by one particle averages.  It is unclear when this diagram commutes.}\label{fig: NS}
\end{figure}
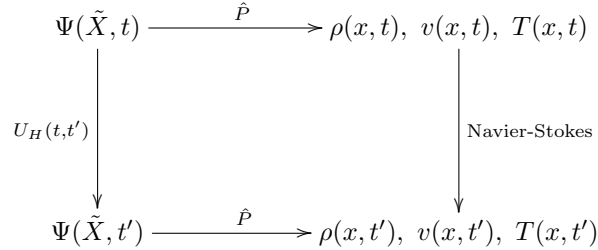
Here we have assumed that $\rho(x)$ is the one body density function defined by integrating out $N-1$ of the coordinates of the many-body density.  The definition of $v(x)$ is chosen to satisfy the one-body conservation law, $\partial_{t}\rho=-\nabla\cdot(\rho v)$. This will, in general, lead to a displeasing nonlocal definition via the Helmholtz theorem that is not simply related to the phase of $\Psi$ and may contain rotational components, so not correspond to any one-body wavefunction.  The generation of $\rho(x)$ is ultimately a nonlocal action as well but it has the merit of being simpler.  The variable $T$, of course, does not have general meaning even for classical systems.  If we can define a generally smoothed density and velocity profile we can interpret the rest of the excitation energy as thermal.  By the ansatz (const)$\cdot k_{B}T= E_{th}$ we can define such a variable that is consistent with the very high temperature thermalized case.  

Discussions of the GP equation often utilize that one way to think of it is as the product of many one-body wavefunctions (with some Jastrow-like correction that is not rapidly changing as the function evolves).  If this reasoning is also valid for the wavefunction of a gas in the range of higher temperatures then we would instead have to consider the commutation of actions as in Fig.\ \ref{fig: NS2}
where we assume that dynamics enforce a coarse grained view of 
\begin{align}\label{productpsi}
\Psi(\tilde{X})\approx \prod_{i}^{N}\psi(x_{i})
\end{align}
The map $P$ is assumed to give $\rho=|\psi|^{2}$ and $v=\nabla \varphi$ where $\varphi$ is the local phase of $\psi$.  We are hoping a general $\Psi$ will tend to such a state after some period of relaxation.  It is clear there will be many that do not since arbitrary energy eigenstates exist as do few state superpositions of them.  The wavefunction of classical gases we expect does not have so much ``stiffness'' that constrains its possible motion yet there must be some dynamical effect of interactions akin to those that drive thermalization to produce long range order implied by Eqn.\ \ref{productpsi}.

The above picture is appealing at some level but if we seek to derive N-S from this we immediately encounter two problems.  First, the thermal motion must be hidden in the fine scale motions of $\Psi$ that are not part of the macroscopic flow.  Secondly, vorticity exists in classical flows but appears only in a singular fashion in wavefunctions.  Furthermore, we can have very small vorticity densities in fluids, densities that are much lower than a single quantum vortex would give if granularly spread out on the same scale as the interparticle separation.  We will need to update our above mapping $\hat{P}$, specifically its typical inverses, to consider such more general flows.  

%Proposition: Dictionary translation for wavefunction for a classical gas and flow.
Given a $\rho(x), v(x)$ we can extract $|\psi|=\sqrt{\rho}$ and irrotational part for the phase $v=-\frac{\hbar}{m}\nabla \varphi$ by the Helmholtz theorem.  The rotational part of $v^{(r)}=v+\frac{\hbar}{m}\nabla\varphi$ must correspond to the coarse grained average of 3N-1 dimensional vortex lines.  Self consistency dictates these can never end but must go to infinity or close in loops.  Fortunately, the identity $\nabla\cdot\nabla\times v=0$ ensures that this can be satisfied.  We may have to consider the low density wavefunction tails outside the support of our classical distribution of matter to do this but we consider this small scale to be resolvable and just a limitation of our best classical description versus a hindrance to a solution.  We insert these vortex line function in the product function fashion if Eqn.\ \ref{productpsi}, however, we do not assume there will be $N$ of them.  

{\bf Gas Hydrodynamic Wavefunction} (GHW):\\
Our trial wavefunction corresponding to a classical hydrodynamic flow $\rho(x), v(x)$ is: 
\begin{align}
\label{fluidpsi}
\Psi=\hat{\mathcal{S}}_{\{x_{j}\}}\prod_{i}^{N}\psi_{i}(x_{i})
\end{align}
where $\mathcal{S}$ is the appropriate symmetrization operator over the coordinate labels and $\psi_{i}=\psi_{0}+\psi^{(r)}_{i}$ is a sum of our best fit irrotational $\psi$ and a similar $\psi$ with a vortex curve and a damping of near it defined by some characteristic length $\xi$.  We have no immediate ansatz to assign winding numbers so we assume that typically the phase advanced by $2\pi$ around each.  The net effect of all these is to generate the observe vorticity in $v^{(r)}$.  A first guess for the characteristic length $\xi$ would be the thermal wavelength.  However, we have only incorporated bulk flow and density information into our trial wavefunction.  To incorporate thermal information, we note that, in this ``geometric'' local scattering limit, a Maxwell-Boltzmann distribution is necessary if the current arrives is fluctuating parcels much smaller than the mfp.  This lets us define a local distribution of small scale oscillations such that $\psi_{T}(x)\approx \int e^{i\phi_{E}}e^{i k_{T}(E) x}dE$ where $k_{T}(E)$ is taken from the classical M-B distribution and $e^{i\phi_{E}}$ is a set of random phases.  There is a natural cutoff for the integral so that $k(E)>\lambda_{mfp}$.  Our final trial ``typical'' wavefunctions are labelled $\Psi(\rho(x),v(x),T(x))$ is as in eqn.\ \ref{fluidpsi} where each $\psi_{i}(x)$ now incorporates the irrotational and rotational flows as before, $\xi=\lambda_{th}$, and $T(x)$ is embedded in the fine oscillatory structure of each.  Furthermore, we randomize the final product phases as in eqn.\ \ref{psith}.  

Although this is hardly a trivial mapping, it seems to be the simplest way to create a plausible typical $\Psi$ corresponding to a classical flowing gas.  The evolution of this then needs to be shown consistent with N-S for long times which implicitly includes the persistence of the wavefunction in such a form.  Conversely, we want to know that given almost any sufficiently energetic $\Psi$ contained in a volume that it will tend to evolve to such a more limited class of typical thermal and hydrodynamic wavefunctions for long times.  

In this trial (GHW) picture notice that the vorticity enters in a very different way than proposed for a rotating solid in Sec.\ \ref{convection}.  The vorticity now penetrates the support of the wavefunction in an unavoidable way and we have favored products of one-body vortices; a condition that deserves future scrutiny.  Despite its limitations it does have an advantage over the quantum Boltzmann equation.  The derivation of such master equations have occupied tremendous journal space and yet they still seem far from realizing their promise of providing a sound basis for statistical mechanics \cite{Vacchini}.  In terms of a justification for hydrodynamics, the use of such an equation would be a delicate proposition.  The complex part of the wavefunction is encoded in the off-diagonal components of the density matrix $\rho_{ij}$.  These also encode \textit{all} the angular momentum information of the gas.  If we don't keep very carful track of these components this will not be conserved.  The attenuated features of the wavefunction norm that exists at the locations of singular vorticty and its mobility is also rather delicate and it is not clear how such information would be mined from the density matrix.  N-S hydrodynamics is generally viewed as a statement of energy and momentum conservation and is often derived as such \cite{Chorin}.  For a quantum based treatment it seems advantageous to utilize such conserved and topological quantities as its foundation.  We will pursue this as a derivation of the N-S equations from consideration of the wavefunction.  

Given the extreme delocalization of realistic gases it is hard to see how the classical billiard ball picture of gases has much of anything to say about gases that are given by many body wavefunctions.  The utility of simple picture of classical kinetics has endured a long time and its results are consistent with experiment of high temperature gases however, if the delocalization time of a gas is so short, why should it be relevant?  We had some success at a plausible explanation of thermal behavior for a broad class of wavefunctions but
the derivation of N-S from such equations looks even more onerous.  What can we say about the high energy limit to suggest that such a coarse grained ($n/|\nabla n|,~|v|/|\nabla v|\gg n^{-1/3}$) product exists?  

The free body wavefunction is not constrained and maintains broad long range kinematic freedom.  Assuming the typical wavelengths are small compared to $n^{-1/3}$, the interacting case can be thought of of as having $\sim N^{2}$ scattering 3N-3 dimensional rays that emanate from the origin acting the currents of the 3N dimensional wavefunction.  The vast number of these suggests that, when the oscillating wavelengths and currents are very small they will tend to move parallel to them.  This however, is exactly what we expect for a wavefunction whose phase is a product function $\Phi(\tilde{X})\approx \prod \varphi_{i}(x_{i})$ since the scattering from the center at $x_{j}\approx x_{k}$ is only from transverse components to the many body current $\tilde{J}$ in the $(x_{j},x_{k})$ subspace. Let us seek a more general global phase in 3N-D to accomplish this.  Fix the phase on a one body projection $\phi(x)=\Phi|_{x_{2},\ldots x_{N}}$, and seek symmetry preserving deformations that preserve this and give no transverse fluxes at the scattering centers.  Seek a first order deformation of $\Phi=\prod_{i}\phi(x_{i})$ of the form 
\begin{align}\label{productfluid}
\Phi'=\hat{\mathcal{S}}\prod_{i}(\phi(x_{i})+\epsilon_{i} \zeta_{i}(x_{i}))
\end{align}  
 where all the $\zeta_{i}$ functions are linearly independent of $\phi$ and not all proportional to each other.  We are interested in a solution in the span of this basis that obeys the $\sim N^{2}$ transverse flux conditions.  To get a general solution valid on the scale of our coarse graining $n^{-1/3}$ we need a set of $\{\zeta_{i}\}$ with oscillations on this scale.  This places a bound on the oscillation scale by which we can alter our product function approximation that is finer than we associate with the bulk flow properties our gas.  

There are examples where hydrodynamics ``breaks,'' specifically, when no gradient expansion can be accurate.  This can nullify the problems of convergence of higher expansions from Chapman-Enskog theory \cite{Chapman} \cite{Cohen} by making them irrelevant but it does leave some uncertainty as to how to correct the evolution.  One can further ask if there is a time when thermodynamics itself ``breaks.''  For example, if a gas is in a trap where $|V|/|\nabla V|\lesssim \lambda_{mfp}$ the local velocity distribution may become so distorted and the Knudsen number ($Kn=\lambda_{mfp}/L$) so large that a local definition of thermodynamic variables is not valuable.  
If one has a rapidly expanding gas where the radial velocity approaches or exceeds the thermal velocity of the thermal waves or the outwards mean free path diverges, equilibration itself become quenched.  The currents simply have no time to reach the scattering regions about the two-body diagonals to exchange hyperradial  with hyperangular motion.  This suggests that bulk expansion of a fluid may introduce some special problems and corrections that are not of a classical nature.  One could similarly label this situation as being one where the ``classicality'' of the gas (as exhibited by Eqn.\ \ref{fluidpsi}) ``breaks.''  As an example, consider high frequency sound of a packet with width $d<n^{-1/3}$ or $d<\lambda_{mfp}$.  In classical kinetics such a density fluctuation makes no sense.  However, such high frequency contributions are completely allowed in a wavefunction.  On these scales the expansion is governed by unequilibrated motion or the quantum pressure.  Such contributions might provide a physical way to extend the hydrodynamic gradient expansions without convergence problems.  

There are many persistent oscillatory high energy states that are not similar to product functions on any scale.  For the case of contact potentials this is demonstrated with the hyperradial expansion flows in Castin and Werner \cite{Castin}.  Given an $\mathcal{L}^{2}$ wavefunction with discrete symmetry about exchanges of coordinate labels in a harmonic trap, many periodic solutions can be observed even for states that are very far from symmetrized products.  The contact potentials give an exceptional simplicity to the problem that mirrors the free case gas but with boundary conditions at the two-body diagonals.\footnote{Werner \cite{Werner:2009} in his thesis noted that for contact interactions, at higher energies, almost all states are ``noninteracting.''  My opinion is that the limit of a contact potential has eliminated the very process of thermalization. The finite width of the potentials and the shorter wavelength of the fluctuation's oscillations is what gives the lateral changes in motion that cause equilibration.  Similarly, in the case of classical kinetic configurations, they equilibrate in 2D and higher but fail in 1D. }  However, for any system made from spherically symmetric two-body potentials we can equivalently consider this to be just an elaborate external potential with a discrete symmetry under coordinate pair exchanges.  This immediately implies that any wavefunction with similar symmetry will undergo hyperradial expansion that leaves this symmetry preserved.  As such there is no ``damping'' that drives this energy to hyperangular motions that would allow the one body density function $\rho(x)$ to relax and so interpret the evolution as governed by the N-S equations.  These counterexamples shows that the hydrodynamic limit must be somewhat subtle.  The harmonic potential is very special in that it is the only one that gives hyperspherically symmetric isopotentials.  For gases that start from equilibrium, don't endure extreme expansions and are perturbed by forces that have the 3D nature of classical matter we will see that N-S gives a believable evolution for GHW initial data.

\subsection{Navier-Stokes Equations}

Ideally we would like to know how a general energetic wavefunction settles down to something we would recognize as thermodynamic with a well defined 3D notion of density, velocity and temperature.  Even in the classical case, this has been a huge undertaking.  A well accepted treatment of the Boltzmann equation was only recently derived \cite{Villani}.  Therefore we accept a more modest set of goals.  
Given the GHW approximation we need to know 1.\ why such a function tends to arise and 2.\ why it persists.  The previous section discussed the favorability and durability of the the high temperature coarse grained product Eqn.\ \ref{fluidpsi}.  The evolution of such a state thus amounts to finding the evolution of $\rho(x)$, $v(x)$ and $T(x)$.  Since the density is low we know that the currents from the fluctuations carry true momentum and transport mass.  This makes them suitable for deriving real forces and stresses on the system.  The vorticity penetration cost is considered to be negligible so that it can be transported and created in any locally conserved manner necessary to preserve the other conservation laws.

The velocity field $v(x)$ is assumed to have small gradients on the scale of the mean free path $|v|/|\nabla v|\ll \lambda_{mfp}$ and the currents from the thermal velocity transport mass much faster than velocity $v$ changes.  This can be expressed in terms of the collision time $\tau=\lambda_{mfp}/v_{th}$ as $v_{th}\gg\dot{v}\tau$.  Since the evolution equations will determine $\dot{v}$ this is a self consistency condition.  

The internal stress of a given flow is given by the pressure $P$ that measures the rate of reflected momentum in a parcel several mean free paths or larger.  The shear stress is due to the transfer of momentum across mfp sized regions that then average momenta nonlocally.  Since this always results in a decrease in macroscopic kinetic energy this results in internal heating.  Based on previous arguments, MB statistics for the currents hold so the net stress is identical, to this order, with the classical kinetic arguments $\Pi_{ij}=-P\delta_{ij}-2\eta(v_{ij}-\frac{1}{3}\nabla\cdot v)$ where $P$ and $\eta$ are determined by the internal distribution or the thermal velocity and density of the gas \cite{Jeans}.  

Using that these 3D dynamic variables stay well defined for such conditions we can impose momentum conservation to immediately derive \cite{Chorin} 
\begin{align}\label{NSeqn}
\rho(\partial_{t}v+v\cdot\nabla v)=\nabla\cdot\Pi
\end{align}
Together with the conservation of mass condition. $\partial_{t}\rho+\nabla\cdot(\rho v)=0$, we have the usual Navier-Stokes equations.  

This does not seem very impressive.  Essentially, we have done a lot of work to show that the usual hydrodynamic equations hold, at one point, invoking parts of the usual classical kinetic arguments.  A virtue of this is that it reminds us that old arguments with, later revealed, unphysical starting points can still give correct results.  At least as importantly, it gives us a new starting point for looking for higher order corrections.  The usual Chapman-Enskog expansion has serious and long standing problems \cite{Cohen}.  Despite this much labor has been put into deriving corrections from classical virial expansions and molecular dynamics (MD) simulations.  A typical wavefunction approach reveals that higher order corrections can have nonclassical variations that include small scale quantum pressure driven waves and variations from the GHW form.  Inspired by Eqn.\ \ref{productfluid} and the 
bifurcation of the order parameter near the edges of condensed bose gases in Eqn.\ \ref{eqn:rot} we might seek corrections of the form
\begin{align}
\Psi'=\hat{\mathcal{S}}\prod_{i}(\psi_{i}(x_{i})+\epsilon_{i} \zeta_{i}(x_{i}))
\end{align} 
where vorticity is distributed among the, otherwise identical, $\psi_{i}$ and there are a distribution of small irrotational and vorticity corrections included by the $\zeta_{i}$ functions.  
This wavefunction approach suggests that we might be better off modifying the Chapman-Enskog expansion by allowing a multiplicity of $\rho,\,v,\,T$ variables at each point.  The classical kinetic approach is limited to the effect of deformed local velocity distributions and their fluctuations.  A quantum approach allows for the simultaneous superposition of such variations.  

In this discussion we have been primarily interested in the two extremes of high temperature (``classical'') and low internal energy (unavoidably quantum) to lend support for the notion that a single wavefunction approach has credibility.  Now we move on to the implications of this idea for ultracold gases and history dependent effects that imply thermodynamic and hydrodynamic treatments are overreaching.

\subsection{Equilibrium Limit in Gaussian Traps}
We have already seen that there are some difficulties in deriving N-S for gases described by wavefunctions at energies that would ostensibly be ``classical'' even for large particle number.  The problems arise from finding a class of functions that can be adequately mapped onto the usual hydrodynamic variables $(\rho(x), v(x), T(x))$ and showing this mapping remains valid under propagation by the Schr\"{o}dinger equation and that N-S equations give a good coarse grained description in terms of them.  However there is a more basic equilibrium problem that is very germane to the types of experiments we do with ultracold traps in gaussian shaped laser traps.

Consider the case of $N$ classical (billiard-like) particles in such a well.  The potential for such a trap can be well approximated by $V(r)=V_{0}(1-e^{-a^{2}/r^{2}})$ where $r$ is the single particle radial coordinate and the net potential is the many body sum of these.  
If the mean free path is small compared to potential gradients, $\lambda_{mfp}(x) \ll V/|\nabla V|$, we expect MB statistics to be well reproduced locally.  The trap depth is $V_{0}$ and the thermal energy per particle is $E_{th}=\frac{3}{2}k_{B}T$.  If $V_{0}\gg E_{th}$ we can expect an exponential-like damping in the density with height.  However, the MB distribution always has some probability of particles near the edges of the trap (where \textit{outwards} mean free paths diverge) have escape energy.  This gives losses, assuming ergodity, that will continue until the remaining particles have less net energy than the energy for a single particle to escape $N_{r}E'_{th}\lesssim V_{0}$.  Since the energy lost by each escaping particle is $E >
 E_{th}(t)$, we see that the large $N$ case gives almost all particles ejected and the remaining ones having almost zero energy.  Similar losses must exist for self bound gravitational bodies like the sun although with very long lifetimes.  However, in the case of gaussian traps, the forces are much more localized and the escape times much shorter.  

In contrast, a many body wavefunction localized in the trap by cooling must be primarily composed of components that are truly bound states.  Unbound components ($E> 0$) will leak out with characteristic time $\tau\sim\hbar E$.  This leaves the only important loss processes as three body recombination and heating from external sources which are independent and somewhat controllable.  The contrast between these two situations is not directly one of hydrodynamic or thermodynamic behavior but of the necessary existence of a bound stationary state in one case and the practical absence of one in the other.

\subsection{Ultracold Gases}\label{ultracold}
In the case of ultracold bosons, we have already seen that the energy cost of a vortex penetrating the support of the wavefunction can be prohibitive and that surface waves generated by external vorticity can be favorable.  This is especially true for the case of interacting bosons where strong curvature near the two body diagonals makes further attenuation of the amplitude there more expensive and the current's energy, $E_{j}$, must increase to avoid these centers.  The decorrelation of such surface waves as in Eqn.\ \ref{eqn:rot} leads to a state where even the pure irrotational motion cannot be well defined by the classical hydrodynamic variables.  These examples give clearly nonhydrodynamic behavior.   

The temperature of such gases using the microcanonical ensemble would we derived by many-body eigenstates near a given energy $E_{0}$.  Many-body superpositions give currents and fluctuations.  A gas that is evaporatively cooled by reducing the trap depth  allows these fluctuations or higher frequencies to exit the gas.  This causes not just the average energy, $\Braket{E}$ to decrease but also the spread in the energy, $\Delta E$, of the component eigenstates that make up the state.  Magnetic field sweeps alter the interaction of the particles and helps keep the hyperradial directions populated with energy for further evaporation.  Since superpositions give time dependent fluctuations and these always attain local energies greater than the mean $\braket{E}$, this drives the energy spread to zero faster than the mean energy itself.  

For these reasons, it seems fair to assign an evaporatively cooled cloud a well defined temperature.  When $\Delta E/\braket{E}\approx 0$ the microcanonical definition of temperature suggests we assign $T^{-1}=dS/d\braket{E}$ where $S$ is the logarithm of the number of states in window about $\braket{E}$.  However, this does not mean that such a state has any power to equilibrate other clouds or objects to such a state.  If we combine two clouds with the same particle number at different such temperatures corresponding to $E_{1}$ and $E_{2}$, we end up with a cloud with energy $E_{1}+E_{2}$ but $\Delta E\sim |E_{1}-E_{2}|$.\footnote{For a thermalized state we expect fluctuations to make up a large part of the kinetic energy so that $E_{j}\sim E_{s}$ or $\Delta E\gtrsim \braket{E}$ where local oscillations vary much faster than external potential, so such combinations might lead to thermalization. For $T(x)$ to be locally defined by the the nearly free Hamiltonian we still need a kind of coarse product function structure to the system which is not always obviously true.  }

Now let us compare with the result of a less gentle transformation of the wavefunction.  If we make slow enough changes in the potential and interactions the Gell-Mann Lowe theorem guarantees we stay in the same distribution of states we started with and can quasi-statically return to the original state by such a process \cite{Peskin}.  A common way of heating such systems is by abrupt trap release and recapture \cite{Thomas:2005}.  This can be done in a large step or through many small ones.  The distinguishing feature is that the potential is stepwise rather than smoothly evolving so it cannot remove kinetic energy from the wavefunction.  The different paths describing these two methods of changing the internal energy, evaporative cooling and discontinuous trap release, are illustrated in Fig.\ \ref{commute}.

\begin{figure}
\[
\xymatrix @C=5pc @R=5pc{
\Psi(\tilde{X}) \ar[r]^{\text{evaporation}} \ar[rd]_{\text{evaporation}} &\Psi_{gs}(\tilde{X})   \ar[d]^{\text{rapid expansion}}\\
&\Psi^{*}(\tilde{X})}
\]
\caption{The wavefunction for a gas evaporatively cooled to finite thermal $E_{th}$ versus cooled to the ground state then heated by abrupt release to reach the same thermal energy.  The diagram does not commute. }\label{commute}
\end{figure}
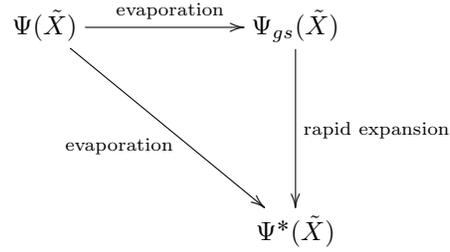

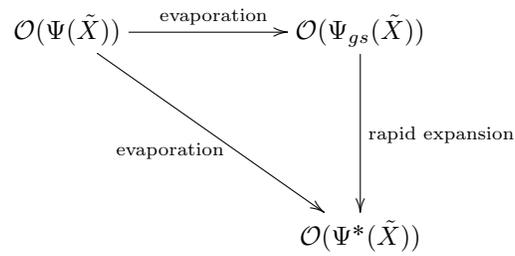
\begin{figure}
\[
\xymatrix @C=5pc @R=5pc{
\mathcal{O}(\Psi(\tilde{X})) \ar[r]^{\text{evaporation}} \ar[rd]_{\text{evaporation}} &\mathcal{O}(\Psi_{gs}(\tilde{X}))   \ar[d]^{\text{rapid expansion}}\\
&\mathcal{O}(\Psi^{*}(\tilde{X}))}
\]
\caption{Some observables of a many body wavefunction e.g.\ the cloud radius, may commute under this action if they are only a function of the net energy.  }
\end{figure}\label{commute1}

If thermodynamic equilibrium is obtained we would expect all observables of the system to be independent of the history of its preparation as in Fig.\ \ref{commute1}.  
The observables of a cloud in a spherical trap include its radius $R$ and details of its density $\rho(x)$.  The velocity is not directly observable but vortex motion can be detected by characteristic density variations and sometimes by careful interference of the matter waves.  Momentum distributions are calculated from free expansion of the cloud.  The density function $\rho(x,t)$ can be observed and we can measure depth averaged values of $\dot{\rho}$.  Perpendicular measurements typically allow an accurate assessment of $\dot{\rho}$ at every point.  
If this was a product function we could extract $v$ from the Helmholtz theorem assuming a local conservation law.  The resulting many body momentum distribution is then easily found from that of a single state and the multiplicity of copies of it.  
If the interaction was slowly turned off before the expansion and the system was initially in the ground state, this would be accurate.  In general, it is not entirely clear that a one-body velocity function $v(x)$ is well defined.  For it to be meaningful, we would expect a local hydrodynamic set of equations to close as a function of $v$ and information derived from $\rho(x)$.  For the general many-body wavefunction a histogram of $\Re \Psi^{\star}\tilde{\nabla}^{2}\Psi$ gives a well defined measure of the distribution of kinetic energy but it is unclear how this relates to something we can measure.  We also are often more concerned with one-body kinetic contributions $\Re \Psi^{\star}{\nabla}^{2}_{x_{k}}\Psi$ but for interacting or highly correlated functions it is also not clear how to extract the distribution of these values from data either.  Presumably, scattering experiments could yield some velocity information but I am unaware of if this is an idea that has fruitfully progressed.  Other measurable quantities are fluctuations \cite{Meineke} \cite{Esteve}, three body recombination rates and tunneling rates from a trap.  

For a narrow $\Delta E$ state (near eigenstate) above the grounds state, $E_{j}/E_{s}\rightarrow0$.  This tells us that the tunneling rates from a gaussian trap should vanish and the three body recombination should be at a minimum.  The relaxed cloud radius $R=\braket{r^{2}}^{1/2}$ should be largely insensitive to the size of $\Delta E$ since it is a measure of the balance between the net kinetic and potential energy.  However, the details of the attenuation of the cloud we would expect to be different.  Higher frequency waves in the hyperradial direction will extend further from the trap center.  To quantify this, let $\Delta E$ become large enough so that the typical eigenstates of energy $E_{0}$ have a radius $R(E_{0})$ that exhibits a nonlinear behavior over the range $E_{0}\pm\Delta E$.  Specifically, if $R(E_{0}+\epsilon)\approx R_{0}+A(\epsilon-E_{0})+B(\epsilon-E_{0})^{2}$ then we expect a difference in the difference in the trap radius $\Delta R= R_{heating}(E_{0})-R_{cooling}(E_{0})\approx \frac{1}{12}B \Delta E$.  

Interestingly, the most accurate measurements of the presumed equation of state (EoS) have been done during cooling \cite{Ku:2012}.  It would be interesting to make a systematic study of the variation in experimental results based on cloud preparation and see if they correlate with history dependence from heating that introduces such internal persistent currents.   

To enhance our ability to probe tunneling loss rates we can introduce a barrier potential by replacing the usual gaussian shaped laser beam with a cylindrical varying beam that generates a barrier potential of height $V_{0}$, with resonant states, $V(r)=V_{0}(\alpha e^{-a^{2}/r^{2}}-e^{-a^{2}/(r-b)^{2}}-e^{-a^{2}/(r+b)^{2}})$
This is most easily realized in 2D with a (not necessarily narrow) harmonic confinement in the z-direction.  An advantageous property about this potential is that we can start with our usual gaussian trap and produce the the desired $\braket{E}$ and $\Delta E$ by the usual methods, then optically convert to this new potential.  By adjusting $\braket{E}-V_{0}\gtrsim\Delta E$ we can observe tunneling losses out of the trap at a rate much faster than three body recombination.  

So far we have only discussed variations in the internal structure due to radial changes in the external potential.  There are two other experimental handles at our disposal: interaction strength and angular momentum.  The possibilities for angular momentum to distribute itself were partially discussed in Sec.\ \ref{angular}.  Starting from a rotating elliptically deformed cloud of interacting bosons in a spherical trap the one-body function $\rho(x)$ will relax to an ellipsoidal deformed cloud.  Hydrodynamics predicts a final rigid body cloud rotation but the strong internal curvature can raise the energy cost of vortex penetration to large for the initial angular momentum present to enter.  The barrier for this is lowered if the interaction strength is lowered or the thermal energy is increased.  Each of these should make a change in the shape of the clouds, rate of evaporative losses and the ratio of angular momentum to mass loss carried with this flux of evaporated atoms.  

The case of fermionic gases is special in several ways.  Firstly, there is no natural reason for vorticity to correlate in them leading to some visible depression in the one-body density function $\rho(x)$.  This leads to the opinion that they are not ``superfluid.''  Additionally, we can control the range of interaction over a much larger range while keeping the clouds stable thanks to the effect of antisymmetry that keeps amplitude low at the n-body diagonals where amplitude can get transferred to small bound states that leave the trap.  Three body recombination is the dominant such process.  

So far we have made extensive arguments that thermodynamics and hydrodynamics do not properly apply to bosonic ultracold gases.  Fermionic gases are more complicated but similar arguments apply although the consideration of the distribution of angular momentum is evidently more complicated.  For free bosons near their ground state in a trap we can have oscillations persist indefinitely and, due to Madelung reformulation of Schr\"{o}dinger dynamics, these are hydrodynamic.  Bosonic clouds with interactions exhibiting small oscillations also obey an Euler equation approximation for the evolution.  
For free fermions, a similar situation exists but it does not make itself evident in the one-body density function $\rho(x)$ due to the many frequencies that must compose it due to the antisymmetry of same spin label coordinates.  

It is however very interesting that near unitarity, $a_{sc}k_{F}=0$, hydrodynamic evolution seems to be recovered.  The Euler equation gives ``elliptic flow'' in free expansion rather than the ballistic motion we would see for classical kinetics at low density or free fermion waves.  Small oscillations correspondingly give long lifetimes of coherent motion for $\rho(x)$ which would rapidly become an undifferentiated static ellipsoid in the free case.  
The antisymmetry of the wavefunction drives the wavefunction to zero for all the $\mathcal{O}(N/2)^{2}$ same spin label diagonals.  In a spin unpolarized gas, there are many more diagonals where the function gets driven to zero, specifically, all the pairs of coordinates where the spin labels are $(\uparrow,\downarrow)$.  It seems that this is sufficient to drive the cloud to a restricted set of velocity fields and a stiffness of evolution so that the motion of $\rho(x)$ seems to closely track the case of the cloud under quasi-static multipolar shifts in the potential.  We gave a similar argument for why interacting bosonic clouds do this Sec.\ \ref{quantumgas}.  In the case of bosons the interactions are strongly repulsive (else the cloud collapses).  For fermions near unitarity we have attractive interactions that are just at the threshold for bound states to occur.  If the interaction is too strong we have bound pairs that repel (due to the effect of symmetry that is still evident from their composite fermion structure).  This seems not to be enough to enforce hydrodynamic behavior as these pairs get smaller.  The threshold case gives long scattering lengths, $a_{sc}>n^{-1/3}$, so that the ground state curvature can enforce correlations that strongly constrain the evolution.  It would be interesting to prepare a state with large $\Delta E$ so that the time varying currents are comparable to this $E_{s}$.  Presumably, this would destroy hydrodynamic behavior faster for clouds with the same net energy and the same strength of interactions.

The question of damping is an important one.  In hydrodynamics, it measure the time for oscillations and nonrigid rotation to settle down to a minimum energy state and is parameterized by the viscosity.  Viscosity and vorticity are closely interrelated in that the final states tend to have uniform vorticity and even purely irrotational flows tend to pull in vorticity at the boundaries as a result of viscosity.  Incompressible fluids without viscosity  subject to conservative forces leave vorticity unchanged.  In the case of our gas clouds, we must first ask what exactly is damping.  Since we now have reason to believe in a history dependent structure hidden in the apparently stationary relaxed cloud density $\rho(x)$ we must ask what observable is damping.  The most evident observable is the cloud density $\rho(x)$ itself.  

Quantum treatments of damping typically begin with linear response theory, generally the Kubo formula.  This can be thought of as a scattering based approach.  It is interesting to consider the same problem from the standpoint of eigenstate superposition.  As we argued for thermalization, some systems will tend to a local universality of current distributions and fluctuations that depend only on the energy density despite being very different distributions of eigenstates.  Relaxation to such states is what we measure in damping.  Because we expect this to be universal behavior that is independent of the eigenstate distributions it seems generally fruitless to approach it from such a point of view.  However, for our cold gases that contain history dependent features this is not necessarily true.   It is entirely possible that knowing the energy distribution of states with a small set of additional information about them that we could derive damping rates in terms of fundamental constants.  

The density of states $g(E)$ is the function on which quantum thermostatistics is built.  For a cloud in a fixed trap we can, more generally,  consider the set of eigenstates partitioned based on other parameters such as mean radius $R$, angular momentum $L$, ellipticity $\eta$ and so forth.  By constructing typical initial data based on distributions of, not just the energy, but these other variables we can consider the damping rates of measurable features such as the distortion of the cloud, ${\eta}$, and the mean radial profile $\braket{\rho(r)}$.  This seems to be the natural extension of thermo and hydro to trapped gases.  As the distributions get larger we may be able to detect  a crossover and convergence to classical hydro and, for the first time, measure a quantum to classical hydrodynamic transition that clearly conserves angular momentum.  

\subsection{Harmonic Oscillations}

There are variety of trapping configurations for holding ultracold gases.  Laser and magnetic fields exploit the low field seeking property of some hyperfine states.  Some of these are now done on chip sets which allow compactness of the device and manipulation features that are novel.  However for optimal isolation and trap shape control optical lattices now dominate the subject.  Although these beams typically have a gaussian profile, the lowest few percentiles of them are well described by harmonic potentials.  This shape induces a high dimensional symmetry that allows a great simplification in the structure of the eigenspectrum.  The gaussian traps were convenient for our discussion above where evaporation and internal currents were our main interest.  However, the many-body shape of the cloud in such a trap is not simple.  The isopotentials exhibit bulging away from the origin in the manner of the $L^{p}$ spaces for $p>1$ of analysis \cite{Rudin}.  This makes the analysis of small oscillations very difficult and, for large $N$, will eventually create tight $n$-body ``corners'' where the curvature of the wavefunction cannot penetrate.

In contrast, the harmonic potential is exceptionally simple.  
To see why this is so consider that our 3N dimensional wavefunction feels an external potential given by $\sum_{i}^{N}V_{ext}(x_{i})$.  For a harmonic (and anisotropic) potential we see that its equipotentials are perfect hyperspheres.  For contact potentials we can view the Hamiltonian as a free Hamiltonian over a space with modified boundary conditions along the two body diagonals.  This has been exploited by Castin and Warner \cite{Castin} to demonstrate that the eigenstates of the free and unitarity bounded systems form towers of states separated by $\Delta \omega=2\omega_{0}$ where the ground state $E_{g'}$ of each tower varies.  Further, they demonstrate that any eigenstate of the trap, when suddenly released or the trap begins to periodically oscillate obey a mathematically simple dynamic behavior \cite{Stringari:2004}.  

For small radial cloud oscillations, this is often taken as proof of zero bulk viscosity in a unitary gas but it is good to reflect on this from the point of view of arbitrary superpositions and how this reflects on observables.  We can choose some linear combination of the ground state and first excited state of a tower and obtain persistent $2\omega_{0}$ oscillations.  However the magnitude of such oscillations is very small since all N particles share in the meager $2\omega_{0}\hbar$ energy.  This is clearly below the threshold of an observable change in the denisty of our cloud $\rho(x)$.  To get large energy oscillations we can take simple superpositions of the ground state and states with $E\sim N\hbar\omega_{0}$.  These states are high frequency and therefore still low in amplitude.  To generate the kind of relatively large amplitude ($\delta R\sim R_{trap}$) low frequency ($\omega\sim\omega_{0}$) motions we observe in experiment let us first consider how it works in the free particle case.  

We can form the states $\psi_{2n}(x_{i})+\alpha\psi_{2n+1}(x_{i})$ for each coordinate where these are one-body states separated by energy $\epsilon=2\hbar\omega_{0}$.  If we do this over all states $\psi_{n}$ up to twice the Fermi level and take an antisymmetrized product we will have our observables persistently oscillate at $\omega'=\epsilon/\hbar$.  Expanding we see that we have a very large range of energy of the many-body eigenstates involved in the superposition.  This distribution will be sharply peaked but the spread is important in obtaining oscillations that are measurably large and of low frequency.  
This introduces a time scale $\tau=\hbar/\Delta E$ from which we can find a damping rate for one-body observables.  However, we must be careful in this interpretation.  If I have a one dimensional wavefunction as the states in a square well and take a superposition of eigenstates with spread $\Delta E$ corresponding to a peak displaced from the center, then the center-of-mass never settles down to an equilibrium value.  
The ``equilibration'' of one-body observables evidently depends on the high dimensionality of the system.  This is familiar from the case of classical kinetics where the 1D case of a gas of hard spheres may never relax.  

An example just discussed \cite{Castin} is the pure hyperradial oscillation of an interacting fermi gas in a harmonic trap never relaxes.  
If we choose our superposition to come from one tower of states (identified by the ground state $E_{g'}$) we have only hyperradial changes to the wavefunction and thus, even though we have some energy spread of eigenstates, the cloud width is a superposition of a one-parameter set of oscillations.  Additionally, experience with interacting bosonic gases \cite{Polkovnikov} show that equilibration in such cases does not happen.  Therefore the actual superpositions involved for damped states must involve some change in the hyperangular motion as well.  This gives the beginnings of what could be a condition on wavefunction superposition for damping of few body observables.  The most important observation is that observable distortions of the the cloud will involve distributions of eigenstates over a broad energy scale and relaxation of some observables will be related to this spread of energy.  

In the hydrodynamic picture of damping, viscosity provides the mechanism and we obtain
\begin{align}
\rho\frac{v}{\tau}\approx \eta \frac{v}{l^{2}}
\end{align}
where $\tau$ is the characteristic damping time and $v$ and $l$ are typical velocities and length scales of the flow.  From this we see $\tau\sim \rho l^{2}/\eta$.  A dimensional observation that $[\eta/\rho]=[\hbar/m]$ suggests a way to relate the damping rate to the quantum of action.  Such arguments are similar to reasoning for why viscosity has a quantum bound.  If one is more skeptical that hydrodynamics is a suitable model for such a system, one can still arrive at such a relaxation time scale for oscillations of the cloud by using the above observation on the energy spread of the distribution of eigenstates involved at unitarity.  

Using that the universal parameter for a unitary gas $\beta\sim\mathcal{O}(1)$, we can say 
the energy spread per particle of $\Psi$ is $\Delta E\sim \frac{\hbar^{2}}{m L^{3}} l$ where $L$ is the cloud size and $l$ the deformation width of the trap at the beginning of the oscillation.  Assuming that there are many hyperangular excitations involved in this superposition, so that it is not a 1D subset, as in the hyperradial expansion case, we have a relaxation time for the trap anisotropy of $\tau\sim \hbar/\Delta E\sim \frac{\hbar}{E}\frac{L}{l}$.  This gives a 
quantum expression of damping that is independent of hydrodynamics.  As noted above, this is the time for the one-body density function $\rho(x)$ to settle down in the trap.  It does not imply that the internal currents and evaporation rates are the same as the cloud before it was released or the same as an evaporatively cooled cloud with the same energy.   

\section{Conclusions}

In the preceding paper we discussed the problems in mapping classical objects on to wavefunctions.  The kinematic freedom of 3D solids and fluids are so restrictive compared to the general many body wavefunction that it is not clear when this is possible.  The issue is somewhat confused by solid state treatments of the electron portion of a solid's wavefunction that exhibits the symmetry and discreteness we imagine of classical bodies.  True grounds states exhibit a rotational delocalization we don't observe in nature and is not present in these solutions because these functions are only functions of the electron coordinates.  By generating a specific wavefunction for solids with phonon excitations we have a sufficiently clear example, along with gases with high internal energy and short range interactions, to consider thermalization and the paradoxes that surround attempts to view equilibration in terms of a single wavefunction's evolution.  By establishing a local condition on equilibration we arrive at a definition of temperature that depends on the energy of the system but not on the net eigenstate distribution of the system and a required modification of the thermodynamic limit.  This allows a notion of thermodynamics that is sufficiently universal for high energy systems and allows very cold systems, as in ultracold gas physics, to retain history dependent features while still acting sufficiently hydrodynamic and exhibiting some apparent transient relaxation while not being truly thermodynamic.  

Gases are distinct from solids in that they cannot maintain the long lasting localization of bound large mass objects.  This requires that any consideration of hydrodynamic evolution consider delocalization and reconcile it with apparent 3D behavior.  This introduced some additional energy costs to the presence of vorticity that solids did not have.  The appearance of a 3D description for high energy gases is not the result of wavefunction ``stiffness'' as is common for very low energy gases, that is one-particle language are referred to as degenerate, but from the minimization of long range scattering in flows.  This limit is rather delicate and it is not clear how fast arbitrary initial data will tend to it.  
It seems that higher corrections to hydrodynamic behavior may have more to do with failure of the system's ``classicality'' resulting in multivalued descriptors than higher terms in the classical Boltzmann expansion.  

Ultracold gases and their dynamical behavior has motivated much of the hydrodynamic treatments of these systems.  In some cases, these have been fairly successful.  In other cases, data seems inconsistent with various theories and experiments.  Our treatment in terms of single wavefunction evolution (pure states), instead of with density matrices or hydrodynamics, predicts history dependent effects in the evaporation, recombination rates and profile as well as some small changes in the cloud extent.  Damping is discussed as a possible measure of relaxation of particular one-body observables rather than viscous hydrodynamics.

The author gratefully acknowledges conversations with Thomas Sch\"{a}fer, Dean Lee and Lubos Mitas.

\bibliographystyle{plain} %Formats bibliography
\cleardoublepage
\normalbaselines %Fixes spacing of bibliography
\clearpage
%\addcontentsline{toc}{chapter}{Bibliography} %adds Bibliography to your table of contents
\bibliography{References} %your bibliography file - change the path if needed
%-----------------------------------------------------------------------------%
\endgroup

\end{document}